\documentclass[acmsmall, nonacm ,screen]{acmart}

\AtBeginDocument{%
  \providecommand\BibTeX{{%
    Bib\TeX}}}

\usepackage{pifont}

\usepackage[ruled,vlined]{algorithm2e}
\usepackage{enumerate}
\usepackage{subcaption}
\usepackage{listings}

\setcopyright{none}

\begin{document}

\title{Handling out-of-order input arrival in CEP engines on the edge combining optimistic, pessimistic and lazy evaluation}

\author{Styliani Kyrama}
\authornote{Both authors contributed equally to this research.}
\email{kyrastyl@csd.auth.gr}
\orcid{0000-0002-5592-8506}
\author{Anastasios Gounaris}
\authornotemark[1]
\email{gounaria@csd.auth.gr}
\orcid{https://orcid.org/0000-0003-3976-7615}
\affiliation{%
  \institution{Aristotle University of Thessaloniki}
  \city{Thessaloniki}
  \country{Greece}
}

\renewcommand{\shortauthors}{Kyrama and Gounaris}

\begin{abstract}
  In Complex Event Processing, handling out-of-order, late, and duplicate events is critical for real-time analytics, especially on resource-constrained devices that process heterogeneous data from multiple sources. We present LimeCEP, a hybrid CEP approach that combines lazy evaluation, buffering, and speculative processing to efficiently handle data inconsistencies while supporting multi-pattern detection under relaxed semantics. LimeCEP integrates Kafka for efficient message ordering, retention, and duplicate elimination, and offers configurable strategies to trade off between accuracy, latency, and resource consumption. Compared to state-of-the-art systems like SASE and FlinkCEP, LimeCEP achieves up to six orders of magnitude lower latency, with up to 10 times lower memory usage and 6 times lower CPU utilization, while maintaining near-perfect precision and recall under high-disorder input streams, making it well-suited for non-cloud deployments.
\end{abstract}




\maketitle

\section{Introduction}

Modern applications, including telemedicine and remote patient monitoring (RPM), require systems capable of handling data from multiple sources, must meet several requirements including scalability, real-time responsiveness, secure and private processing, fault tolerance, stateful data management, and reliable storage \cite{hartmann2022edge}. These requirements are often addressed through hybrid architectures that integrate edge and cloud computing paradigms \cite{dimitriadis2022scalable, su2021cloud, nikolov2023container,prabhu2022edge, dilibal2020development}
where data is generated at the edge by IoT devices, preprocessed near the edge, and forwarded to the cloud for further analysis. This hierarchy allows close-to-source processing, filtering irrelevant data, and preserving only useful insights \cite{hurbungs2021fog}. Processing data closer to sources also addresses privacy concerns by limiting the exposure of data to public networks, especially in the case of medical applications, where sensitive information is included in the patient's data. Local, on-device processing enables immediate processing of data, thereby reducing both communication latency and processing delays associated with cloud-based systems.
One of the most useful rationales for low-latency stream processing is Complex Event Processing (CEP) \cite{cugola2012processing, dayarathna2018recent}. Our work is motivated by practical problems when deploying CEP solutions locally in resource-constrained environments, as detailed below.

\textbf{Motivation Scenarios.}
Consider an RPM system for patients with chronic conditions or recovering from major surgeries. Integrating a CEP system into RPM enables real-time detection of events of interest, improving decision-making based on patient health data \cite{dautov2019hierarchical, lan2019universal}. A deployment setup may include \texttt{location sensors} for tracking movements, a \texttt{smartwatch} for monitoring daily activity and heart rate, a \texttt{smart-scale} for weight measurements, and a \texttt{smart vest} for real-time vital sign monitoring, such as heart rate, RR intervals, and accelerometer data. 

These sensors produce data at varying rates: location sensors emit updates instantly during transitions or periodically when the patient is stationary, the smartwatch transmits data every minute, the scale sends data immediately upon use, usually once per day or once per week, while the smart vest provides continuous updates every second on patient’s vital signs. Processing such heterogeneous data streams in distributed settings requires the system to be resilient to network instability, sensor failures, and data inconsistencies, such as delays, duplicates, or disorders. These inconsistencies can significantly impact the quality of results (QoR) of a streaming application that requires real-time output \cite{aquacep}. Therefore, ensuring low latency while responding to these issues is crucial for maintaining quality.

For instance, the RPM system could detect an impending anxiety crisis, indicated by ten minutes of immobility combined with a sudden spike in steps recorded by the smartwatch, such as in \autoref{q:q1}, or recognize early signs of a cardiac event by correlating rising heart rate data from the smart vest with increased sweat levels from the smartwatch, as defined in \autoref{q:q2}.

\begin{lstlisting}[label=q:q1,caption=Impending anxiety crisis, captionpos=b,frame=tb!]
PATTERN SEQ(!ROOM a, STEPS+ b[])
WHERE b[i+1].value > b[i].value
WITHIN 10 minutes
\end{lstlisting}

\begin{lstlisting}[label=q:q2,caption=Early signs of a cardiac event, captionpos=b,frame=tb!]
PATTERN SEQ(HR+ a[], SWEAT b)
WHERE a[i].value < a[i+1].value 
AND b.levelIncreased == true
WITHIN 5 minutes
\end{lstlisting}

Out-of-order, duplicate, batch, or even missing data can result from faulty sensors due to incorrect time settings, poor network connectivity, or other functional issues. When sensor data arrives late, out of order, or in batches (e.g., as may happen with smartwatch data post-patient absence), it may lead to misdetections or missed events, hence affecting the quality of system’s results and providing misleading monitoring about the patient. The application of this setup is not restricted to patient monitoring.

\begin{figure}[tb!]
    \centering
    \includegraphics[width=0.7\linewidth]{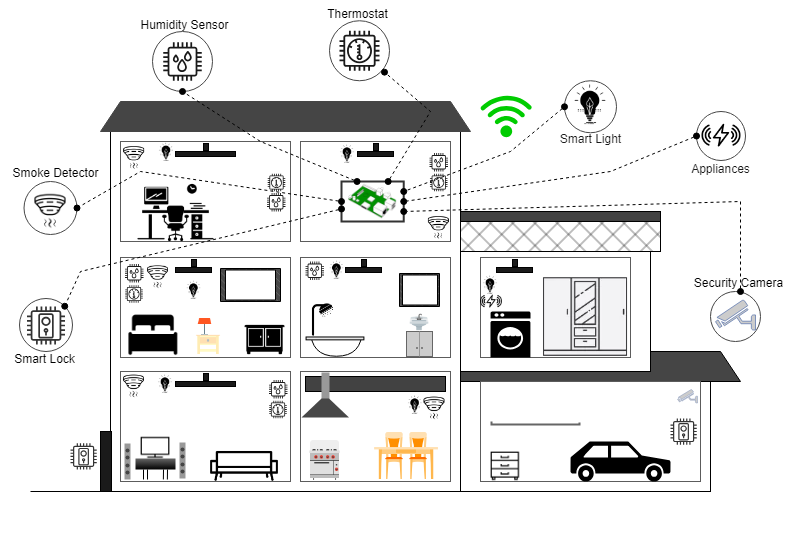}
    \Description{This image shows a Smart-Home setting including all sensors}
    \caption{A Smart-Home setting including all sensors}
    \label{fig:Smart-Home scenario setting}
\end{figure}

Similar challenges arise in domains like smart home automation, particularly for managing security and emergency response. A smart home may integrate devices such as temperature and humidity sensors, smart locks, security cameras, smoke and gas detectors, and motion sensors, all connected to a central edge-processing hub, as depicted in \autoref{fig:Smart-Home scenario setting}. In such a setting, a CEP-enabled edge system processes data locally to enable both privacy and real-time responsiveness. For instance, detecting a fire may require correlating gas leak alerts, rising temperature readings, and smoke detection within a 30-second window. This enables prompt identification and immediate response to prevent escalation. Such a pattern can be represented with the following query \autoref{q:q-smarthome}\footnote{thresholdGas and thresholdSmoke could be either user-defined or set by domain experts}:

\begin{lstlisting}[label=q:q-smarthome,caption=Q-Smart Home,captionpos=b,frame=tb] 
PATTERN SEQ( GasLeakSensor a, TemperatureSensor+ b[], SmokeDetector c ) 
WHERE a.percentage > thresholdGas
AND b[i+1].temperature > b[i].temperature 
AND c.percentage >= thresholdSmoke
WITHIN 30 seconds 
\end{lstlisting}

Order-sensitive processing is vital; incorrect event sequences (e.g., smoke detected before gas may suggest a harmless situation occurring in everyday life, like cooking smoke, rather than a fire risk) could mislead emergency detection. Efficient handling of such inconsistencies ensures accurate, timely responses even in the presence of sensor delays or failures.

\textbf{Summary of Novelty and Contribution.} 
This work addresses the challenges of CEP in resource-constrained environments, focusing not only on minimizing latency but also on balancing the trade-offs between accuracy, timeliness, and computational cost in the presence of real-world data inconsistencies, such as out-of-order, late, and duplicate events.

Advanced CEP techniques include lazy processing \cite{kolchinsky2015lazy, corecep, cetpoppe}, buffer-based mechanisms \cite{ji2015quality, ji2016quality, zacheilas2017maximizing, weiss2020dynamic, mutschler2013distributed} and speculative processing \cite{barga2006cedr, liu2009sequence, li2007event, brito2008speculative,mutschler2013reliable, mutschler2014adaptive}, however, addressing only one aspect, and even when dealing with inconsistencies, most solutions are not resource-aware.
To address these limitations, we introduce LimeCEP, which combines the above approaches in a novel manner, exploiting the advantages of each technique, and resulting in a hybrid solution. To avoid re-inventing the wheel, LimeCEP also leverages the functionalities offered by Apache Kafka with regards to message ordering, retention, and duplicate removal features.

Our contribution is five-fold:
\begin{enumerate}[i.]
    \item We propose a novel hybrid approach that integrates lazy processing, buffer-based mechanisms, and speculative techniques. This combination enables efficient handling of real-world data inconsistencies, such as late, out-of-order, and duplicate event arrivals, while ensuring the detection of maximal matches.
    \item LimeCEP investigates the trade-offs between accuracy and latency in CEP systems combined with computational cost, providing a configurable framework for handling late and duplicate arrivals. In this way, we aim to optimize both timeliness and quality of results.
    \item Our approach is designed to support both a relaxed strategy and a non-deterministic relaxed one, as supported by most CEP systems, employing optionally correction of emitted matches, while supporting both single-query and multi-query detection, ensuring applicability to multiple scenarios.
    \item LimeCEP is designed for deployment in resource-constrained environments, by utilizing shared structures, suitable for deployment on resource-constrained environments, such as edge devices and non-cloud infrastructures, where traditional CEP systems, such as FlinkCEP and SASE, are impractical due to their high resource demands. In our evaluation, while SASE grows up to 6.5 GB and FlinkCEP up to 8 GB of memory with increased CPU usage up to 90\% in complex scenarios,  LimeCEP remains highly efficient, requiring only 1.2 GB for the most demanding case, and just 50–300 MB in smaller or simpler setups, with a CPU usage up to 15\% even in the most demanding cases.
    \item We provide an extensive experimental evaluation of our approach, demonstrating the applicability to real-world scenarios where inconsistencies occur, and the efficiency in maintaining a balance between accuracy and latency, while reducing computational cost. Our evaluation shows that LimeCEP consistently outperforms state-of-the-art systems such as SASE, SASEXT, and FlinkCEP. It maintains near-perfect precision and recall under high-disorder input, and achieves up to six orders of magnitude lower latency.
\end{enumerate}

LimeCEP can be downloaded in open source format from \url{https://github.com/KyraStyl/LimeCEP} 

The rest of this paper is structured as follows. Section \ref{sec:rw} reviews related work in the field of CEP, highlighting existing limitations. Next, Section \ref{sec:notation} introduces the notation and formal definitions used throughout this work, along with a clear statement of the problem. The proposed solution is detailed in Section \ref{sec:solution}, which covers the high-level design, architectural components, adaptations for multi-query detection, and the complete algorithm, while the handling of duplicate events is discussed in Section \ref{sec:duplicates}. Our experimental evaluation is presented in Section \ref{sec:evaluation}. The paper concludes with Section \ref{sec:concl}, which summarizes our findings and outlines directions for future work.

\section{Related Work}
\label{sec:rw}

CEP has been an active research area, with many systems implemented to identify meaningful patterns over event streams. One of the most representative systems is SASE\cite{sase}, a prototype that integrates a declarative query language with an execution engine that operates incrementally by utilizing NFAs to detect patterns. Even though it performs well under optimal situations, SASE makes assumptions about in-order input and lacks mechanisms for dealing with real-world inconsistencies like disorders, missing events, or duplicates.

FlinkCEP, the CEP library of the Apache Flink framework \cite{flink}, is a newer, high-throughput system that can be applied to distributed environments. It introduces more advanced mechanisms  for disorder handling based on watermarking and punctuation, to allow some out-of-order input tolerance. FlinkCEP mostly targets cloud infrastructures and thus is not the optimal solution for deployment in resource-constrained environments due to its high computational and memory requirements. Furthermore, punctuation and watermarking techniques require the data source to actively participate in disorder management, since punctuations must be explicitly generated by the data source to indicate progress in event time. This dependency limits flexibility.

Numerous improvements on architectures as well as new rationales have been proposed to address performance limitations and ensure result quality under real-world conditions.
CORE\cite{corecep} adopts an indexing technique to enhance pattern detection and reduce reliance on automata. 
Similarly, SASEXT\cite{sasext} is an extension to SASE in which NFAs are replaced with hash-structures, and for complex patterns—such as those involving Kleene+ events—it restricts computation to maximal matches, inspired by the rationale proposed in \cite{cetpoppe}.
All of these changes are made to limit memory overhead and computation complexity while improving latency and throughput.
Many systems adopt lazy evaluation like SASEXT; for instance, OpenCEP\cite{kolchinsky2015lazy} uses lazy evaluation methods to delay computation until triggering events are available for match construction, reducing intermediate state and redundant processing. This helps to reduce resource overload but can increase latency in some cases.

Another group of research works aim to enhance quality and robustness in the occurrence of noisy input. For example, Mutschler et al. \cite{mutschler2013reliable, mutschler2014adaptive} suggested speculative processing with k-slack buffering to manage out-of-order event arrival in distributed event-based systems. They dynamically adjust the level of speculation according to CPU utilization by wrapping the event detector with a speculative reordering layer that delays or replays events based on dynamic slack estimation. The detector is treated as a black box and is assumed to receive in-order input. To recover from mis-speculations, the system is considered to follow a snapshot-based rollback mechanism. While this rationale may be very effective in pub/sub systems, this decoupled design (detector layer-speculative layer) is not fully compatible with existing CEP systems, such as SASE and FlinkCEP. CEP systems mostly rely on incremental evaluation over internal stateful match structures like NFAs or indexed partial matches. In such systems, disorder affects not only input order but also in-progress partial matches, requiring internal mechanisms for reprocessing, match invalidation, and retraction. The approach of \cite{mutschler2014adaptive} does not address how a CEP engine will restore the internal state of the NFA to a previous state, re-evaluate active matches, or process replayed events. For this reason, although our solutions are partially inspired by the work in \cite{mutschler2013reliable, mutschler2014adaptive}, we do not use it as a competitor in our evaluation.

Additionally, Rivetti et al.\cite{rivetti2018probabilistic} proposed a probabilistic model, namely ProbSlack, for CEP engines to dynamically resize window bounds and delay evaluation if needed, for each incoming event. This prediction model tries to reduce the limitations of fixed-size slack solutions, yet it is based on the modeling of per-event network distributions. ProbSlack \cite{rivetti2018probabilistic}, which is implemented as an extension over Apache Flink framework\cite{flink}, operates as a scheduling layer that dynamically delays window evaluation to balance completeness and latency, leveraging FlinkCEP for pattern detection. It assumes that, by delaying the evaluation long enough, disorders will be mitigated, but does not handle  the re-evaluation of partial matches or match retraction within the CEP engine in case of late arrivals like LimeCEP does.

LimeCEP combines in a novel way the three main evaluation approaches: slack-based buffering (pessimistic), speculative processing (optimistic), and lazy evaluation, while also computing only maximal matches in the presence of Kleene+ events. The proposed framework provides robust and effective processing of real-world input streams that may contain unordered and duplicated events. Furthermore, our approach avoids traditional NFA-based representations altogether and employs an index-based internal model suitable for resource-constrained environment. It minimizes redundant computation through selective re-processing and on-the-fly statistics to further guide adaptive behavior. This unified design supports high accuracy, low latency, and low resource utilization, outperforming state-of-the-art paradigms, such as FlinkCEP.

\section{Notation \& Problem Statement}
\label{sec:notation}

In this section, we introduce the notations and definitions, along with the problem targeted. The formal definitions of the terms and symbols used are summarized in \autoref{tab:key-symbols} and \autoref{tab:formalisms}. 

\begin{table}[tb!]
\centering
\caption{Frequently used notation and interpretation}
\label{tab:key-symbols}
\footnotesize
\def\arraystretch{1.2}
\begin{tabular}{p{3cm}p{10cm}}
\toprule
\textbf{Notation} & \textbf{Description} \\ \midrule
$e$ & An event instance. \\
$et$ & Event type. \\
$t_{\text{gen}}$ & Timestamp of event generation. \\
$t_{\text{arr}}$ & Timestamp of event arrival. \\
$s_{et}$ & A source of data for type $et$.\\
$S$ & All sources. \\
$P$ & A pattern query, defined by event types, time window, and constraints. \\
$E$ & All set of event types. \\
$E_p$ & The set of event types in the pattern P. \\
$\sigma$ & The pattern structure (e.g., $\text{SEQ}$, $\text{AND}$, $\text{OR}$, repetition operators) \\
$W_p$ & The time window for the pattern $P$. \\
$Q_p$ & Constraints or predicates associated with the pattern. \\
$S_p$ & The selection policy of pattern $P$ (either STNM or STAM). \\
$endT_p$ & The last event type $et$ in $P$. \\
$lastEndT_p$ & The last event $e$ arrived for which $e.et == endT_p$. \\
$I_p$ & Index structure for incoming events for pattern $P$, mapping event types to TreeSets. \\
$TS_{et}$ & A TreeSet containing events of type $et$, sorted by $t_{\text{gen}}$. \\
$STS$ & Shared Treeset Structure. A special type of $I_p$ for multi-pattern scenario. \\
$M$ & A match for a pattern query. \\
$M_{\text{max}}$ & A maximal match for a pattern query. \\
$LM$ & List of Matches $M$. \\
$LM_{\text{max}}$ & List of Maximal Matches $M_{\text{max}}$. \\
$lta$ & Latest $t_{\text{gen}}$ arrived at the system. \\
$\text{OOO}(e)$ & Out-of-order score of an event. \\
$slc$ & Slack duration. \\ 
$MPW$ & Maximum potential window. \\
$\theta$ & Late event threshold, for discarding late events. \\
$\theta_{s_{\text{et}}}$ & Late event threshold for a specific event type et. \\
EM & Event Manager component. \\
RM & Result Manager component. \\
SM & Statistical Manager component. \\
$ne_{\text{all}}$ & Total number of events processed by the system. \\
$ne_{et}$ & Total number of events of type $et$ processed by the system. \\
$no_{\text{all}}$ & Total number of out-of-order events processed by the system. \\
$no_{et}$ & Total number of events out-of-order events of type $et$ processed by the system. \\
$Q_{\text{to-events}}(P)$ & Events of interest. Mapping of patterns to event types. \\
$E_{\text{to-patterns}}(et)$ & Related patterns. Mapping of event types to patterns. \\ 
$T$ & Kafka topic. \\
$p_i$ & i-th partition of $T$. \\
$ms_{i,j}$ & j-th Kafka message in $p_i$.  \\
\bottomrule
\end{tabular}
\end{table}

\begin{table}[tb!]
\centering
\caption{Definitions of key terms}
\label{tab:formalisms}
\footnotesize
\resizebox{\linewidth}{!}{%
\begin{tabular}{p{3.7cm}p{9.5cm}}
\toprule
\textbf{Term} & \textbf{Definition} \\ \midrule
$e$ (Event) & 
$e = (id, et, t_{\text{gen}}, t_{\text{arr}}, s_{et}, payload)$, where: \\
& \begin{tabular}[t]{@{}l@{}}
$payload$: Additional attributes associated with the event, e.g. ``value''. \\
Example: e = (1, Steps, 10:15:30, 10:20:10, Fitbit, value = 125)
\end{tabular} \\ \hline

$P$ (Pattern) & 
$P = (E_p, \sigma, S_p, W_p, Q_p)$, where: \\
& \begin{tabular}[t]{@{}l@{}}
$E_p$, such that $E_p \in E$, under $S_p$ selection policy. \\
Example: P = ($E_p =\{A,B,C\}$, $\sigma=\{A+,(B$ or $C), D\}$, $S_p$=STNM, \\ $W_p=10$ minutes, $Q_p=\{A[I].value>A[i-1].value\}$)\\
\end{tabular} \\ \hline

$lastEndT_p$ (Last endEvent) &
$lastEndT_p =  TS_{endT_p}.getLastEvent()$ \\ 
& \begin{tabular}[t]{@{}l@{}} 
Example: if $\sigma = \{A,B,C\}$ and $TS_{C} = \{C_2, C_5\}$\\ 
then, $lastEndT_p = C_5$
\end{tabular} \\ \hline

$I_p$ (Index Structure) & 
$I_p = \{ et \rightarrow TS_{et}\} \forall E_p$\\ \hline

$STS$ (Shared Treeset Structure) &
$STS = \{ et \rightarrow TS_{et}\} \forall E$ \\ 
& \begin{tabular}[t]{@{}l@{}} 
Example: if $E = \{A,B,C,D,E\}$, then \\
$STS = \{A\rightarrow TS_A, B\rightarrow TS_B, C\rightarrow TS_C, D\rightarrow TS_D, E\rightarrow TS_E\}$
\end{tabular} \\
\hline

$M$ (Match) & 
$M = \{e_1, e_2, \ldots, e_k\}$, such that: \\
& \begin{tabular}[t]{@{}l@{}}
i. $e_i \prec e_{i+1}$, i.e. $e_i.t_{\text{gen}} < e_{i+1}.t_{\text{gen}}$ ,  $\forall i \in [1,k$), \\
ii. Events {$e_1, e_2, \ldots, e_k$} satisfy $Q_p$, \\
iii. $e_k.t_{\text{gen}} - e_1.t_{\text{gen}} \leq W$.
\end{tabular} \\ \hline

$LM$ (List of Matches)
 & $LM = \{ M_1, M_2, \ldots, M_k\}$ \\ \hline

$c(e,m)$ (Compatible event) & An event $e$ is compatible with a match $M$ for a pattern $P$ denoted by $e \sim M$, when: \\
& \begin{tabular}[t]{@{}l@{}}
i. $e.et \in E_p$, \\
ii. $e$ satisfies $Q_p$, \\
iii. $e.t_{\text{gen}} > e_j.t_{\text{gen}}$ for all $e_j \in M$ \\
then, $e$ can extend match $M$. \\
Example: $P = \{\sigma=SEQ, E_p=\{A,B,C\}, Q_p=\{A.value>B.value\}, W_p=10 \}$ \\
$M=\{A_1\}$ with $A_1.value=5$ and $e=\{et=B, t_{\text{gen}}=2, value=3\}$ \\
since $e.et=B \in E_p$, $e.value=3<A_1.value$, and $e.t_{\text{gen}}-A_1.t_{\text{gen}} \leq W_p$, \\
then $e \sim M$. 
\end{tabular} \\ \hline

$M_{\text{max}}$ (Maximal Match) & 
$M_{\text{max}} = \{e_1, e_2, \ldots, e_k\}$ is maximal for $P$ if:\\
& \begin{tabular}[t]{@{}l@{}}
i. satisfies all constraints $Q_p$, \\
ii. $\nexists e$ such that: $e \sim M_{\text{max}}$ and $M_{\text{max}} \cup \{e\}$ satisfies $Q_p$ 
\end{tabular} \\ \hline

$LM_{\text{max}}$ (List of Maximal Matches)
 & $LM = \{ M_{\text{max}_1}, M_{\text{max}_2}, \ldots, M_{\text{max}_k}\}$ \\ \hline

$\text{OOO}(e)$ (Out-of-Order Score) & 
Out-of-order score for an event $e$, defined as: \\ 
& \begin{tabular}[t]{@{}l@{}}
$ \text{OOO}(e) = log(1+time\_diff) + arrival\_diff^2 + norm\_window\_perc$\\
i. $time\_diff = e.t_{\text{gen}} - \text{latest\_}{t_\text{gen}} $ \\
ii. $arrival\_diff = |estimated\_rate(e.et) - actual\_rate(e.et)|$ \\ 
iii. $norm\_window\_perc = \frac{actual\_rate(e.et)}{window\_length}$
\end{tabular} 
\\ \hline

$aff(e,LM_{\text{max}})$ (Affects prior Maximal Matches) & An event $e$ affects previously computed maximal matches if:
\begin{tabular}[t]{@{}l@{}}
$e.t_{\text{gen}} < lta$ AND  
$(e.et == endT_p$ OR $e.t_{\text{gen}} < lastEndT_p.t_{\text{gen}})$
\end{tabular} \\ \hline

$\theta_{s_{\text{et}}}$ (Late Event Threshold) &  $\theta_{s_{\text{et}}} = 2.5 *$average\_ooo\_score(et) //2.5 is actually a configurable variable\\ \hline

$extl(e)$ (Extremely Late Event) & $extl(e): if (OOO(e) > \theta_{s_{\text{e.et}}})$ \\ & \begin{tabular}[t]{@{}l@{}} An event $e$ is discarded if $extl(e)$.
\end{tabular}\\ \hline

$Q_{\text{to-events}}(P)$ & $Q_{\text{to-events}}(P) = E_p$. \\ \hline

$E_{\text{to-patterns}}(et)$ & $E_{\text{to-patterns}}(et) = \{P_i \mid et \in E_{p_i}\}$. \\

\bottomrule
\end{tabular}
}
\end{table}

\paragraph{\textbf{Selection Policies}}
Besides the conditions $Q_p$ specified in a pattern $P$, selection policy also affects the way relevant events from stream $s_{et}$ are selected to form matches, by implicitly defining contiguity conditions. There are three main and widely known policies:
\begin{enumerate}
    \item \emph{Strict contiguity (SC)}, which allows events to be exclusively consecutive in the input stream, without irrelevant events existing in-between.
    \item \emph{Relaxed contiguity (RC)}, which does not require proximity between the relevant events, but matches each event with the next relevant event found in input stream, ignoring pattern-irrelevant events. It is also referred to as \emph{Skip-Till-Next-Match (STNM)} policy.
    \item \emph{Non-deterministic relaxed contiguity (NDRC)}, which operates greedy upon the selection, by matching each event with every possible upcoming event. This policy is also known as \emph{Skip-Till-Any-Match (STAM)}.
\end{enumerate}

We target two variants of finding pattern matches in event streams under realistic situations, where events may arrive out of order. The simpler one refers to a single query pattern and is defined as follows.

\begin{definition}[Single Query Detection]
Given a pattern $P = (E_p, \sigma, S_p, W_p, Q_p)$, the goal of a Complex Event Processing (CEP) system is to identify matches that satisfy the specified structural $\sigma$, temporal $W_p$, and constraint-based $Q_p$ conditions taking into account both computational efficiency and result completeness. 
\end{definition}

In general, as we have argued, inconsistencies such as late or out-of-order, and duplicate event arrivals introduce significant challenges, affecting both the accuracy and efficiency of CEP systems. 
When a new event $e$ is processed, its inclusion in or exclusion from $LM_{\text{max}}$, the list of maximal matches for $P$, depends on the temporal and structural constraints $W_p$ and $Q_p$. Late or out-of-order events (events for which: $e.t_{\text{gen}} < lta$ \&\& $e.t_{\text{arr}}>= lta$) can impact previously computed maximal matches $LM_{\text{max}}$ if $e.et$ corresponds to $endT_p$ (i.e., their type triggers our lazy evaluation procedure) or if $e.t_{\text{gen}} < lastEndT_p.t_{\text{gen}}$. Inadequate handling of these inconsistencies can lead to a degradation in the QoR.

Existing approaches mainly aim at detecting either all possible matches (e.g., \cite{sase}), ensuring completeness but with high computational and memory overheads, or only maximal matches (e.g., \cite{cetpoppe}), proposing more efficient solutions and reducing resource usage and latency when patterns contain Kleene clauses. We adopt the latter rationale for wider coverage of pattern types, but our novelty focuses on another limitation, that is, most methods in the literature lack adaptive mechanisms to address late or out-of-order, and duplicate events. Although systems like FlinkCEP handle these inconsistencies using watermarking techniques, their high resource consumption makes them unsuitable for resource-constrained environments (e.g., non-cloud infrastructures).

A fundamental challenge in this context is maintaining both high accuracy and Quality of Results (QoR) while minimizing latency. This involves determining the optimal Maximum Potential Window (\emph{MPW}) dynamically, which represents the maximum set of events that must be retrieved and reprocessed to resolve the effects of a late or out-of-order event $e$. An overestimated \emph{MPW} leads to unnecessary computations, increasing latency, while an underestimated \emph{MPW} risks missing relevant matches, thereby degrading the QoR. This tradeoff is particularly pronounced in resource-constrained environments, where achieving a balance between computational efficiency (in terms of number of events that need to be processed and the associated latency) and accuracy is essential for detecting only maximal matches in real-time. To address the above challenges, LimeCEP decides \emph{MPW} in an adaptive manner. 

The definition above, is extended to handle also multiple queries:

\begin{definition}[Multiple Queries Detection]
Given a set of patterns $\{P_1, P_2, \ldots, P_n\}$, the goal of a CEP system is to identify matches for each query in the set taking into account both computational efficiency and result completeness. 
\end{definition}

\section{Proposed solution}
\label{sec:solution}

\subsection{Higher level aspects \& architecture}
\label{sec:architecture}

\begin{figure}[tb!]
    \centering
    \includegraphics[width=0.65\columnwidth]{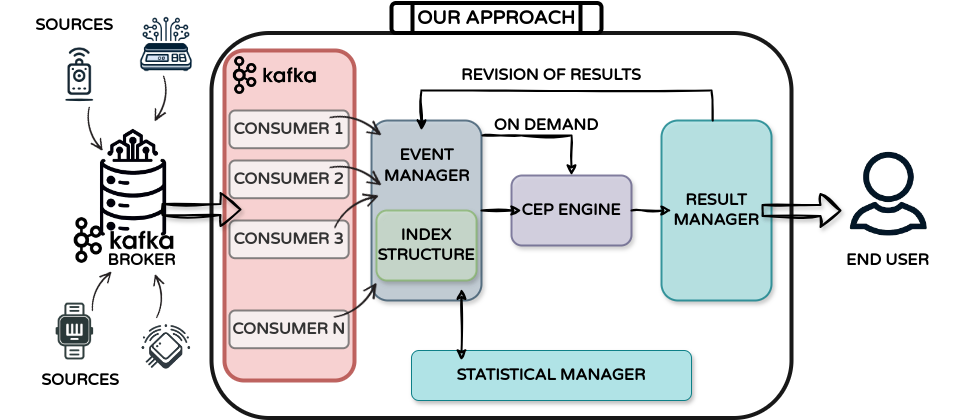}
    \Description{This image shows the Architecture for Single-Query Detection}
    \caption{Architecture for Single-Query Detection}
    \label{fig:architecture}
\end{figure}

The main components of our solution are depicted in \autoref{fig:architecture} and are discussed in detail below.
The system consists of five key components apart from the core CEP engine:
\begin{itemize}
    \item \textbf{Kafka Consumers:} Responsible for collecting incoming data from multiple sources and preprocessing them to transform them into a unified format.
    \item \textbf{Index Structure:} An efficient event storage mechanism utilizing TreeSet for maintaining the total ordering of events.
    \item \textbf{Event Manager (EM):} The central processing component, that filters and processes events before passing them to the CEP engine.
    \item \textbf{Result Manager (RM):} Maintains computed matches and ensures users receive accurate event detection results.
    \item \textbf{Statistical Manager (SM):} Computes and updates key metrics for tracking event ordering, timeliness, and out-of-order statistics.
\end{itemize}

\subsubsection{Kafka Consumers}
For the first layer of data collection, Kafka consumers are employed, to receive all input from multiple topics. From the query pattern $P$, the set of relevant data sources $S$ is extracted, and for each member of this set, there is a distinct Kafka topic. Once the messages arrive, they are parsed and transformed into single or multiple events according to the unified format presented in \autoref{tab:formalisms}.
Messages \(ms_{i,j}\), where \(i\) denotes the partition $p_i$ and \(j\) represents the sequence within the partition, are segmented and processed in parallel, with Kafka guaranteeing \(ms_{i,1} \prec ms_{i,2} \prec \dots\) (total order within \(p_i\)). However, as global ordering across partitions (\(p_1, p_2, \dots, p_n\)) is not guaranteed, inconsistencies may arise when \(ms_{i,j}\) and \(ms_{k,l}\) (for \(i \neq k\)) are processed.

\subsubsection{Index Structure}
\begin{figure}[tb!]
    \centering
    \includegraphics[width=0.2\columnwidth]{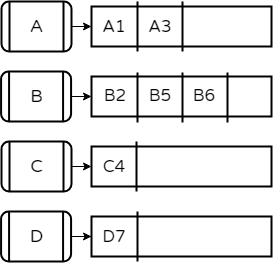}
    \Description{This image shows a Simple TreeSet-based index structure for four sources}
    \caption{Simple TreeSet-based index structure for four sources}
    \label{fig:treeset}
\end{figure}

Incoming events are stored in an indexed structure $I_p$, which contains a treeset $TS_{et}$ for each member of $E_p$. For example, if $S=$ \{$s_A,s_B,s_C,s_D$\} and $E_p=$\{\verb|A, B, C, D|\}, $I_p$ will have four different keys. For each one of these event types $et$ that serve as keys to the hashmap, the corresponding Treeset $TS_{et}$ stores the actual events. The rationale behind choosing TreeSet as the underlying data structure, stems from its inherent  properties that align with the system's requirements. More specifically, TreeSet is a combination of a TreeMap and a SortedSet, therefore, it is simultaneously a self-balancing binary search tree and a sorted collection. The treemap structure provides us with the much needed efficiency for operations such as adding, removing and retrieving elements. Furthermore, the combination of the two results in the organization of events by custom elements, in our case by \( t_{\text{gen}} \), ensuring total ordering, and $O(log(n))$ complexity of the insertion of incoming out-of-order arrivals. In addition, by utilizing the functions \texttt{equals()} and \texttt{hashcode()}, which are implemented in the object class, it is guaranteed that any duplicate input will be discarded from the TreeMap. An example of a simple \emph{Treeset-based Index-Structure} $I_p$ with four sources is shown in Figure~\ref{fig:treeset}. Each event is denoted by a letter indicating the $s_{et}$, and a number indicating $t_\text{gen}$.

For single queries, the index structure is encapsulated in the Event Manager, described below.

\subsubsection{Event Manager}

Event Manager (EM) stands as an intermediate stage between the data collection stage and the processing of events by the CEP Engine. This core component is mainly responsible for (i) storing all incoming data in the appropriate structures, (ii) deciding whether or not the CEP engine should receive these events for processing, and (iii) updating all the necessary statistics for the system to work.

Upon the arrival of an event $e$, the Event Manager (EM) first evaluates its relevance to the pattern $p$ by verifying whether $e.et \in E_p$. If $e$ is not relevant, it is immediately discarded. Otherwise, it proceeds to the timeliness evaluation. Event $e$ is assessed as to whether it arrived in-order or out-of-order; in both cases, it assigns an \emph{out-of-orderness} score, denoted as $OOO(e)$, which quantifies the event’s deviation from the expected order. If $e$ is in-order, $OOO(e) = 0$, otherwise, the score is assigned a positive value, affected by multiple factors described later in this section. 
An event $e$ is stored in the appropriate Treeset $TS_{et}$ of $I_p$ only if it satisfies both relevance and timeliness constraints. 
A \texttt{late event threshold} $\theta$ is calculated by EM per source $s_{et}$. Currently, we employ the formula $\theta_{s_{et}} = \alpha * \text{average\_ooo\_score(et)}$, where we set $\alpha=2.5$, but other formulas, e.g., based on the standard deviation of the \emph{OOO(e)} scores, are possible. Each incoming event $e$ is evaluated upon this threshold $\theta_{s_{et}}$. If $OOO(e) > \theta_{s_{e.et}}$ this event $e$ is considered as \texttt{extremely late}, i.e. $extl(e)$ is true.
In that case, $e$ is discarded despite being relevant, ensuring that only temporally coherent events contribute to the pattern detection process, since results containing these events may not be useful to the end user. However, the event is late but $extl(e)$ is false, then, with the help of $MPW$, we ensure that only a limited set of past events are reprocessed.
Overall, after relevance and timeliness checks, $e$ is evaluated upon whether it should trigger the CEP engine to start finding matches or not.

Most CEP systems pass each event through an NFA, and generate intermediate results that may even not reach the final state, to produce a final match $M$. However, this is both time- and memory-consuming. The proposed system operates in a lazy manner, and invokes the CEP engine 
only when necessary, opposite to NFA-based CEP systems that continuously process and evaluate each incoming event. If all the events arrive in order, the system invokes the CEP engine only in case an end event $l$ has arrived where $l$ is of end type, i.e. $l.et = endT_p$. Otherwise, the incoming events are stored in $I_p$ and the processing is stalled. The CEP engine is also invoked whenever an out-of-order event $e$ that affects the already produced matches arrives at the system (details are provided in Section \ref{sec:details}). 

In addition, inspired by the works of \cite{cetpoppe, core} and loosely coupling our solution with the SASEXT engine, the CEP system calculates only the maximal matches $M_{\text{max}}$. This can make pattern detection more efficient, less time-and memory-consuming, especially when Kleene operators are included in the query pattern. This allows our solution to be deployed in a resource-constrained environment. 

In simple queries, that is pattern queries that contain only simple events without repetition, e.g., $P=(\{A,B,C\}, \sigma=SEQ(\{A,B,C\}), W_p=10\text{ minutes}, Q_p=\{A.value < B.value\}) $, maximal matches $M_{\text{max}}$ and simple matches $M$ are identical, so no information is missed. Matches $M$ are defined as a set of events $\{e_1, e_2, \ldots, e_k\}$ where each event $e_i$ is proceeding of $e_{i+1}$, i.e., $e_1.t_{\text{gen}} < e_2.t_{\text{gen}}$, $e_2.t_{\text{gen}} < e_3.t_{\text{gen}}$, and so on. For these events $\{e_1, e_2, \ldots, e_k\}$, conditions of $Q_p$ are satisfied. Moreover, temporal conditions must be met, meaning the difference between the match's start and end times should not exceed $W_p$. For example, for the events $\{A_1, A_2, B_3, C_4, A_5, B_6, C_7\}$, form the matches $M_1=\{A_1, B_3, C_4\}, M_2=\{A_2, B_3, C_4\}, M_3=\{A_5, B_6, C_7\}$, satisfying also temporal conditions, i.e., $C_4.t_{\text{gen}} - A_1.t_{\text{gen}} \leq W_p$, $C_4.t_{\text{gen}} - A_2.t_{\text{gen}} \leq W_p$, $\&$ $C_7.t_{\text{gen}} - A_5.t_{\text{gen}} \leq W_p$.

However, sets of $M$ and $M_{\text{max}}$ differ in the case of Kleene+ presence, since maximal matches are only a subset of all possible matches, i.e., $M_{\text{max}} \subseteq M$. For example, if the previous pattern contain repetitions and thus becomes a complex query, e.g., $P=(\{A,B,C\}, \sigma=SEQ(\{A+,B,C\}), W_p=10\text{ minutes}, Q_p=\{A.value < B.value\}$, then the maximal matches formed are $M_{\text{max}_1}=\{A_1, A_2, B_3, C_4\}$, $M_{\text{max}_2}=\{A_1, A_2, A_5, B_6, C_7\}$. We argue that maximal matches can provide valuable insights in the presence of Kleene+, and can be effectively utilized to derive all possible matches using a simple post-processing function \cite{cetpoppe}. All the final maximal matches $M_{\text{max}}$ are then forwarded to the Results Manager component.

In both simple and complex patterns, LimeCEP supports only STNM and STAM selection policies, since the first one (SC) is considered very strict for real-world scenarios. However, integrating it in our approach is a straightforward modification. We consider STNM as the default Selection Policy for LimeCEP, due to resource-aware rationale. We do not employ consumption policy on LimeCEP's pattern matching process (consumption policies are described in \cite{giatrakos2020complex}), as neither SASE nor SASEXT use consumption policy to prune the produced matches. Since we are inspired by these systems, and we are loosely coupled with SASEXT, we follow the same direction. However, consumption policies can be applied as a post-processing step, similar to the AfterMatchSkipStrategies\footnote{\url{https://nightlies.apache.org/flink/flink-docs-master/docs/libs/cep/\#after-match-skip-strategy}} applied by FlinkCEP to the results of the pattern detection. More details on what is detected per each selection policy, and what is considered maximal or not, are provided in \footnote{\url{https://github.com/KyraStyl/LimeCEP/blob/master/semantics.md}}

In a nutshell, EM is the orchestrator component for the processing of events for a specific query, since it initializes and handles the CEP Engine, updates the statistics, and handles out-of-order arrival of events.

\subsubsection{Result Manager (RM)} Complementary to EM, there is one component responsible for receiving and maintaining a more organized overview of the matches $M_{\text{max}}$, as per the events contained in each match. RM aims to update the user on valid matches, and thus it performs under two different settings: \texttt{with correction} or \texttt{without correction}, however, considering the first one as the default option due to quality-aware rationale. 
RM tracks match status using three key indices:
\begin{enumerate}
    \item \texttt{emitted}, which indicates whether a match has been emitted to the user, and
    \item \texttt{ooo}, which identifies out-of-order matches caused by late-arriving events.
    \item \texttt{updated}: which reflects whether a match has been corrected after its initial emission.
\end{enumerate} 

Out-of-order matches arise due to late-arriving events, indicating either previously missed matches or matches altered by newly incorporated events.
More specifically, RM maintains all matches indexed by their last event, which facilitates efficient lookup and management operations. Through this indexing, several checks can be achieved effortlessly. These are:

\begin{itemize}
    \item \textbf{validity check}: under STNM policy, RM checks if a newly arrived event invalidates previously emitted matches. For instance, if a match \texttt{a9 b12 c19} was emitted, but a late event \texttt{b11} arrives that forms a more valid match \texttt{a9 b11 c19}. Then, RM will invalidate the prior match, update the internal state, and emit the correct one updating the end user. In STAM policy this does not hold, since each event is matched with all possible consecutive events of the next eventtype.
    \item \textbf{maximality checks and correction}: in case of OOO matches detection, RM determines whether a new match is a \emph{maximal extension} of an existing one. If so, it corrects the original match with the extended one, updates relevant indices and statistics, and propagates the corrected result to the user. For example, the maximal match \texttt{a9 b12 b14 b16 c19} is detected but a late event \texttt{b11} arrives, the new maximal match that will be detected is \texttt{a9 b11 b12 b14 b16 c19}. The previous will be corrected with the addition of \texttt{b11}. 
    \item \textbf{existence checks}: regardless of the selection policy, when a new match arrives, RM checks for identical already emitted matches using the indices. These checks aim to eliminate duplicate output, thereby ensuring consistency and robustness.
\end{itemize}

Additionally, RM computes the detection latency (based on the ingestion time of the first event in the match) and updates the general statistics maintained by the system, specifically by the Statistical Manager component, to accumulate the overall average latency for monitoring purposes. Moreover, RM periodically removes expired matches based on a sliding time window, compacting its internal data structures to maintain efficiency. 

In short, RM ensures users receive updates on valid results and propagates changes when matches are affected by late event arrivals.

\subsubsection{Statistical Manager (SM)}
As already mentioned, during the processing of incoming events and match extraction, useful statistics, necessary for the solution, are maintained by a dedicated component, namely \emph{Statistical Manager}. These measurements are calculated and constantly updated by EM, and they are used for decisions throughout the processing of each event, as they reflect the state of each source. SM maintains general counts such as the total number of events $ne_{all}$ processed by the system, as well as the number of events per source $ne_{s_{et}}$ , and similar counts for the out-of-order events arrived, $no_{all}$ and $no_{s_{et}}$, respectively (\autoref{tab:key-symbols}). Moreover, it contains information provided by the user regarding the estimated arrival rate per each source as $esar_{s_{et}}$, and calculates on-the-fly information regarding the actual arrival rate $acar_{s_{et}}$ based on each incoming event (\autoref{tab:stats}).

Three global metrics are maintained to capture the characteristics of late events: the average ($\text{avg\_ooo}$), maximum ($\text{max\_ooo}$), and minimum ($\text{min\_ooo}$) times that an event can be considered out-of-order. Corresponding source-specific metrics are stored in three hashmaps: $Avg_{s_{et}}$, $Max_{s_{et}}$, and $Min_{s_{et}}$, which track the $avg\_ooo_{s_{et}}$, $max\_ooo_{s_{et}}$, and $min\_ooo_{s_{et}}$ values for each source $s_{et}$, respectively. Additionally, the SM maintains a hashmap, $avg\_ooo\_score_{s_{et}}$, which records the average out-of-order score ($OOO(e)$) for each source $s$. This metric is utilized in calculating the \texttt{late event threshold} $\theta_{s_{et}}$ for source $s_{et}$, thereby enabling adaptive and informed decision-making in response to varying event characteristics.

\begin{table}[tb!]
\centering
\caption{Statistic Manager measurements}
\label{tab:stats}
\begin{tabular}{p{3cm}p{10cm}}
\toprule
\textbf{Notation} & \textbf{Description} \\ \midrule
$esar_{s_{et}}$ & The estimated arrival rate for events of type $et$, generated by source $s_{et}$. \\
$acar_{s_{et}}$ & The estimated arrival rate for events of type $et$, generated by source $s_{et}$. \\
$avg\_ooo$ & Average out-of-order time for an event $e$. \\
$max\_ooo$ & Maximum out-of-order time for an event $e$.\\
$min\_ooo$ & Minimum out-of-order time for an event $e$.\\
$avg\_ooo_{s_{et}}$ & Average out-of-order time for an event $e$ of type $et$.\\
$max\_ooo_{s_{et}}$ & Maximum out-of-order time for an event $e$ of type $et$.\\
$min\_ooo_{s_{et}}$ & Minimum out-of-order time for an event $e$ of type $et$.\\
$Avg_{s_{et}}$ & Hashmap for maintaining $avg\_ooo_{s_{et}}$.\\
$Max_{s_{et}}$ & Hashmap for maintaining $max\_ooo_{s_{et}}$.\\
$Min_{s_{et}}$ & Hashmap for maintaining $min\_ooo_{s_{et}}$.\\
$avg\_ooo\_score_{s_{et}}$ & Hashmap for maintaining average $OOO(e)$ for each source $s_{et}$.\\
\bottomrule
\end{tabular}
\end{table}

\subsection{Modifications for multi-query Detection}
\label{sec:multiple}

In real-world scenarios, having a CEP system to detect only a single pattern is impractical. For instance, in the case of the smart home, multiple patterns may be interesting for detection, in order to prevent an upcoming situation and allow for in-time action. Even more in a patient monitoring scenario, multiple patterns should be detected at the same time, to ensure the instantaneous response from the caregiver, in case of emergency. In such scenarios, deploying the previously described system, without any modification, would necessitate running numerous separate instances, each one responsible for detecting a single pattern. To overcome this limitation, the system has been modified to support simultaneous multi-pattern detection within a single instance, optimizing computational efficiency. In this subsection, we describe in detail the modifications made to the system, to support multiple queries.

\subsubsection{Shared Treeset Structure} The first optimization focuses on improving the index structure used for storing incoming events. 
For example, let us consider the case of having the queries $P_1$ \verb|PATTERN SEQ(A+, B, C)| and $P_2$ \verb|PATTERN SEQ(B, C+, D)|, where event types \verb|B| and \verb|C| appear on both.

A naive approach would be to deploy separate system instances or EMs for each pattern to be detected, however, leading to redundant storage and increased memory usage, as overlapping event types are stored multiple times. For instance, $I_{p_1} = \{ TS_A, TS_B, TS_C\}$, and $I_{p_2} = \{TS_B, TS_C, TS_D\}$, where \verb|B| and \verb|C| events are stored twice. To address this, a \emph{Shared TreeSet Structure} (STS) is proposed that maintains a single Hashmap with Treesets $TS_{et}$ for each unique event type of the incoming events, regardless of the patterns they belong to, i.e. $STS = \{ et \rightarrow TS_{et} \forall E \}$. Therefore, for the above example, $STS = \{ TS_A, TS_B, TS_C, TS_D\}$. 

Each pattern query $P$ has its own time window $W_p$, event ordering, and conditions $Q_p$. Therefore, it needs to be handled separately, and consequently, deploying multiple EMs cannot be avoided. Since $I_p$ is replaced with the shared $STS$, the shared storing structure is no longer part of a single EM instance. Instead, it is elevated to a higher architectural layer, making it accessible to all EM instances, thus offering uniform event handling across the system while maintaining event data efficiently in a single, centralized structure. Therefore, each EM should only contain a list of \texttt{events of interest} $Q_{\text{to-events}}(P) = E_p$, and an inverted mapping of related patterns $E_{\text{to-patterns}}(et) = \{ P_i | et \in E_{p_i}\}$ is created by the system to indicate which EM instances should be updated for each $et$ when a new event $e$ arrives at the system.

\subsubsection{Event Manager Instance} 
The dedicated EM instance created for each pattern $p_i$ acts as a pattern-specific orchestrator within the broader architecture. In multi-pattern version EM instance is similarly responsible for:
\begin{enumerate}
    \item maintaining an internal mapping $Q_{\text{to-events}}(P)$, which links the pattern $P$ to relevant event types $E_p$,
    \item evaluating whether incoming events are relevant for processing by the CEP engine based on the $Q_{\text{to-events}}(P)$, temporal constraints $W_p$, and additional optional conditions $Q_p$,
    \item handling the storage of relevant events in the $STS$,
    \item calculating the $OOO(e)$ score for each event $e$ regarding the pattern $p$ handled, and
    \item updating aggregated system-wide necessary statistics through the \emph{Shared Statistical Manager} component.
\end{enumerate}

By leveraging $STS$ for efficient event storage and deduplication, as well as adopting a lazy evaluation approach, the EM instance minimizes memory usage and computational overhead. Lazy evaluation ensures that matches are computed only when an end event $e$ corresponding to the query pattern arrives ($e.et = endT_p$), or when an out-of-order event requiring reprocessing is detected. Remember that each EM instance computes \texttt{maximal match} $LM_{\text{max}} \subseteq LM$, with regards to the pattern assigned.

\begin{figure}[tb!]
    \centering
    \includegraphics[width=0.8\columnwidth]{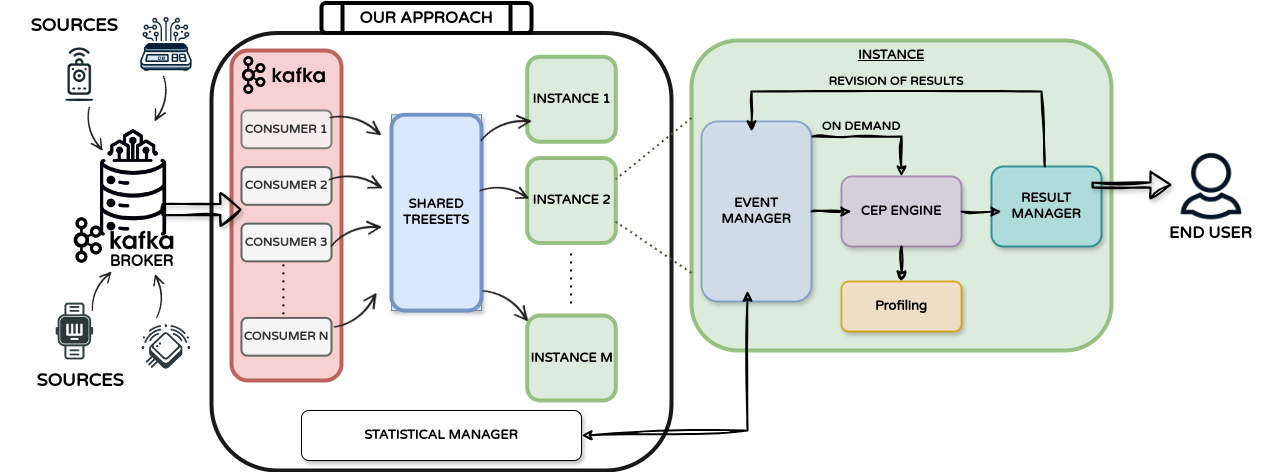}
    \Description{This image shows LimeCEP's architecture for Multi-Pattern Detection}
        \caption{LimeCEP's architecture for Multi-Pattern Detection}
    \label{fig:architecture-multi}
\end{figure}

\subsection{The full algorithm}
\label{sec:details}

In this subsection, an overview of the overall LimeCEP functionality is provided. 
The overall algorithm of our enhanced solution for multi-pattern detection is presented in~\autoref{alg:pseudocode}. Our solution ensures that events are processed efficiently and matches are recalculated accurately while maintaining up-to-date statistical data. By leveraging modular components and shared resources, the solution optimizes event processing and tries to minimize resource utilization.

\begin{algorithm}[htbp]
\footnotesize
\SetAlgoNlRelativeSize{-1} 

\KwData{\texttt{sources} (list of sources), \texttt{consumers} (list of Kafka consumers), \texttt{topics} (list of topics), \texttt{patterns} (list of patterns), \texttt{STS} (HashMap \texttt{<String,Treeset>}), \texttt{EvMs} (list of EventManagers), \texttt{stats} (shared StatisticalManager)}

\KwIn{\\
\begin{itemize}
    \item \texttt{sources} - List of sources to fetch events.
    \item \texttt{consumers} - Kafka consumers subscribed to specific topics.
    \item \texttt{topics} - Kafka topics for event streams.
    \item \texttt{patterns} - Event patterns used by EventManagers.
    \item \texttt{STS} - A mapping of strings to Treeset for storing intermediate event data.
    \item \texttt{EvMs} - List of EventManager instances handling event patterns.
    \item \texttt{stats} - Shared statistical manager to maintain metrics like out-of-order (OOO) scores.
\end{itemize}}

\KwResult{Event processing results with calculated statistics and triggered matches}

\textbf{Initialize EventManagers:}
\ForEach{\texttt{pattern} in \texttt{patterns}}{
    create EventManager \texttt{EM} for \texttt{pattern}\; 
    add \texttt{EM} to \texttt{EvMs}\;
}

\textbf{Initialize Kafka Consumers:}
\ForEach{\texttt{topic} in \texttt{topics}}{
    c = create\_consumer()\;
    add \texttt{c} to \texttt{consumers}\;
}

\SetKwFunction{ProcessEvent}{KafkaConsumer.processEvent}
\SetKwProg{Fn}{Function}{:}{}
\Fn{\ProcessEvent{\texttt{event $e$}}}{
    \If{\texttt{$e$ is valid}}{
        \texttt{map event $e$ to class object} \;
        \texttt{store to $TS_{e.et}$}\;
        \texttt{update \texttt{EM} in \texttt{$E_{\text{to-patterns}}(e.et)$}} \tcp*{That have this type of event in their pattern}
    }
    \Else{
        \texttt{discardEvent($e$)} \tcp*{Handle invalid or duplicate events}
    }
}

\While{\texttt{true}}{
    \texttt{c.processEvent($e$)}  \tcp*{Event e arrived at some consumer c}
    \ForEach{\texttt{EM} in \texttt{EvMs}: if \texttt{EM} in \texttt{$E_{\text{to-patterns}}(e.et)$}}{
        \texttt{$OOO(e)$ = EM.calculateOOOscore(e)}\;
        \texttt{decision = !EM.extl($e$)} \tcp*{$OOO(e) <= \theta_s$}

        \If{\texttt{decision == true}}{
            \If{\texttt{$e.et == endT_p$}}{
                \texttt{results = EM.triggerEngine($e$)} \tcp*{Calculate matches}
                \texttt{EM.updateRM(results)} \tcp*{Update results manager}
            }
            \ElseIf{\texttt{EM.$aff(e, LM_{\text{max}})$}}{
                \texttt{mpw\_start, mpw\_end = EM.calculateMPW(e)}\;
                \If{\texttt{mpw in $STS$}}{
                    \texttt{mpw = EM.retrieveMPWevents(e)} \tcp*{Retrieve from $STS$}
                }\Else{
                    \texttt{c = consumers.getConsumerForTopic(e)}\;
                    \texttt{mpw = c.getEvents(mpw\_start, mpw\_end)} \tcp*{Fetch from Kafka Broker}
                }
                \texttt{results = EM.triggeronDemandEngine(mpw, e)}\;
                \texttt{EM.updateRM(results)} \tcp*{Update results manager}
            }\Else{
                \texttt{wait(stats.averageOOOsource(s))}\;
            }
            \texttt{EM.updateStats(stats, e)}\;
        }
        \Else{
            \texttt{$STS.get(TS_{e.et}).remove(e)$} \tcp*{Evaluate if event should be deleted}
            \texttt{discardEvent($e$)} \tcp*{Ignore event on this EM}
        }
    }
}

\caption{Simplified pseudocode of our approach}
\label{alg:pseudocode}
\end{algorithm}

\emph{\textbf{Initialization.}}
At system startup, dedicated Kafka consumers are instantiated for each data source, ensuring continuous polling for messages from Kafka topics. In our implementation, these consumers operate in a multi-threaded Java environment, enabling receiving events from multiple sources simultaneously.

For each pattern $P_i$ submitted by the user, a corresponding Event Manager instance is created $EM_i$, responsible for handling the processing pipeline for that specific pattern. During initialization, TreeSets $TS_{et}$ are allocated for each event type $et$ referenced in $P_i$, if not already exist. These structures maintain events in total order based on $t_{\text{gen}}$, facilitating efficient retrieval and insertion operations in $O(log(n))$.

\emph{\textbf{Event Processing Flow.}}
All messages $m$ received through the Kafka consumers are transformed into events $e$ and they are forwarded to the subsequent components. Upon receiving an event $e$, this event is immediately stored in the appropriate TreeSet of the $STS$, and all the related statistics are updated by the Statistical Manager (SM). Then, the system evaluates its relevance against the different EMs, by utilizing the $E_{\text{to-patterns}}(et)$ mapping. If $e.et \in E_p$ for an $EM_i$, this EM $EM_i$ is notified for the newly arrived event $e$. Event $e$ is then processed by all relevant EMs, and for each one, it is evaluated against timeliness. Each relevant EM calculates for this event $e$ a different out-of-orderness score $OOO(e)$ to quantify its deviation from the expected order. $OOO(e)$ score is calculated using Equation~\eqref{eq:ooo_score}, the pattern $p$ and its time window $W_p$, and by comparing $e$ against the latest event of the same type within $TS_{e.et}$. We utilize the following formula:

\begin{equation}
OOO(e) = \alpha * \log(1+\text{time\_diff}) + \beta *\text{arrival\_diff}^2 + \gamma * \text{norm\_window\_perc}
\label{eq:ooo_score}
\end{equation}
where:
\begin{align*}
    \text{time\_diff} &= e.t_{\text{gen}} - \text{latest}\_t_{\text{gen}} \\
    \text{arrival\_diff} &= |\text{estimated\_rate}(e.et) - \text{actual\_rate}(e.et)| \footnotemark \\
    \text{norm\_window\_perc} &= \frac{\text{actual\_rate}(e.et)}{\text{window\_length}}
\end{align*}

\footnotetext{Estimated\_rate is defined at the definition of the sources to the system, while actual\_rate is computed throughout the process when an event arrives and is the average time difference between two messages that are generated from the same source.}

The $OOO(e)$ score measures the degree to which an event e deviates from expected stream behavior by combining three main factors: temporal inconsistencies, rate deviation, and lateness impact. The time\_diff captures how far the event’s timestamp diverges from the expected order, the arrival\_diff quantifies the deviation between actual and estimated arrival rates of the event’s source, and the norm\_window\_perc reflects how much the observed delay or frequency matters in relation to the time constraints of the pattern.

The shared Statistical Manager (SM) updates the average lateness of source $s_{et}$ upon the arrival of a new event $e$. Hence, the threshold $\theta_{s_{et}}$, which may be affected by the arrival of $e$ of type $e.et$, is properly adjusted. The threshold $\theta_{s_{et}}$ is determined as follows, whereas, as already mentioned, other formulas can be applied as well:

\begin{equation}
\theta_{s_{et}} = 2.5 \times \text{avg\_ooo\_score}_{s_{et}}
\end{equation} 

By comparing $OOO(e)$ to this $\theta_{s_{et}}$ it is determined whether $extl(e)$ holds or not, and, in turn, it is determined if $e$ is considered for processing or not. Events for which $extl(e)$ holds for a pattern $p_i$ are ignored by $EM_i$, and evaluated for deletion from the corresponding TreeSet of the $STS$. Events that are considered extremely late for all the relevant $EMs$ are immediately removed from the $STS$.

In the case of $EM_i$ processing $e$, a decision on whether or not to trigger the CEP engine $CEP_i$ of $EM_i$ should be made. The CEP engine is triggered in two cases: (i) if event $e$ is an end event, i.e. $e.et==endT_p$ for $p_i$, $CEP_i$ is triggered regardless of the $ooo\_score$ of $e$; (ii)  if $OOO(e)>0$ (i.e. event is out-of-order), then $CEP_i$ is triggered on-demand to re-compute matches from past events, only in case of $e$ affecting the previous matches, i.e. $aff(e,LM_{\text{max}})$ holds. In that case, a set of events is computed and fed to the $CEP_i$ engine.

In both cases, $CEP_i$ processes the extracted event set to compute maximal matches $M_{\text{max}}$, forwarding results to the Result Manager (RM). If matches are affected by late arrivals, RM marks them as \texttt{ooo} matches in the corresponding list, and updates the end user accordingly for the updated matches. For example, if match (a) A1 B4 C5 was detected and emitted previously, and B2 arrived out-of-order, the new match from the on-demand process is (b) A1 B2 C5. In such a  case, RM will update the user for the match (b) and the user can decide whether both are valid, or one of them is obsolete.

\emph{\textbf{Adaptive Slack Mechanism.}}
To complement Kafka’s capabilities, a slack-based mechanism is introduced. By intentionally delaying result processing the system aims to mitigate false positives and negatives. The slack duration $slc$ is dynamically adjusted based on statistical insights, ensuring adaptive event handling. Higher $slc$ is applied when late arrivals are frequent, while minimal or no $slc$ is used when events arrive on time (more details are provided in the next paragraph).

If recomputation is required due to a late arrival, i.e., a relevant event $e$ for which $aff$ holds true, 
the system waits for a slack interval equal to $slc$ before retrieving the necessary events. To minimize the amount of events reconsidered,  these events fall into the Maximum Potential Window ($MPW$). 

\begin{definition}[Maximum Potential Window (MPW)]

The Maximum Potential Window ($MPW$) is the time range within which event $e$ can influence the pattern detection process. It is determined based on event timestamps and the pattern's time window $W_p$ as defined in  \autoref{eq:mpw}.
\end{definition}

\begin{equation}
MPW =
\begin{cases}
[e.ts, \max(e.ts + W_p, lta)], & \text{if } e \text{ is of start type} \\
[e.ts - W_p, e.ts], & \text{if } e \text{ is of end type} \\
[e.ts - W_p + (n_{\text{right}})*t, \max(e.ts + W_p - (n_{\text{left}})*t, lta)], & \text{if } e \text{ is an intermediate event} \\
[e.ts - W_p + kleene\_start(e), e.ts + W_p], & \text{if } e \text{ is a Kleene event}
\end{cases}
\label{eq:mpw}
\end{equation}

where:
\begin{align*}
    n_{\text{right}} &= \text{number of positions to the right of } e \text{ in pattern} \\
    n_{\text{left}} &= \text{number of positions to the left of } e \text{ in pattern} \\
    t &= \text{time offset} \\
    \text{kleene\_start}(e) &= \text{adjustment factor for Kleene events based on their sequence} 
\end{align*}

Intuitively, the $MPW$ defines the window of events that need to be considered when evaluating $e$’s impact on pattern detection. The number of positions either to left or right is mapped into time offset using the $t$ variable based on the time window $W_p$ and the number of events in the pattern sequence $\sigma_p$. Events within $MPW$ are retrieved either from $STS$ or via Kafka’s storage mechanism, depending on availability.

For example, consider the pattern $P = \{ABCD\}$ and a new event $e$ where $e.et=B$ arrived, then $MPW = [e.ts - W_p + 2*t, \max(e.ts + W_p - 1*t, lta)]$ because when looking back from $e$, we need two positions for $C$ and $D$ events, and when looking forward, we need one position before $e$ for the $A$ event. In addition, the time offset $t$ is calculated as $t=W_p/|P|$, which, in the example, is $t=W_p/4$ since $P$ contains 4 events.

\emph{\textbf{Result correctness.}}
To ensure result correctness despite out-of-order arrivals, our system integrates several complementary mechanisms that together increase robustness. This subsection outlines how these mechanisms interact and what guarantees they provide.

As mentioned previously, incoming events are first evaluated against the lateness threshold $\theta_{s_{et}}$. Events classified as \texttt{extl}(e) are discarded. All other events, even if arriving late, are processed by the EM component, potentially triggering recomputation of affected matches by the CEP engine. When a late but acceptable event arrives, the system evaluates whether this event affects previously detected matches. If so, it computes an MPW set of events, to pass to the CEP engine for the recomputation of matches. When the percentage $ratio$ of OOO events reaches a 10\% threshold (this threshold is actually configurable), the system applies a waiting delay before processing further out-of-order arrivals. The slack time is computed as a fraction of the specific pattern's window: $slc = \text{ratio} \times W_p$. This slack time increases the likelihood of batching related late events together, thus avoiding repeated reprocessing. For each end-event within this window, the CEP engine is re-triggered via an \texttt{onDemand} call, in which, only relevant subsets of the stream are used, minimizing overhead while maintaining completeness. Matches produced during on-demand evaluation are passed to the RM component, which ensures result consistency:
\begin{itemize}
    \item If the match is new, it is emitted to the user.
    \item If a structurally identical match already exists, it is skipped.
    \item If a similar match exists but the new one contains more relevant events (i.e., it is maximal $M_{\text{max}}$), the old match is invalidated and replaced, and the end user is updated upon the correction.
\end{itemize}
This mechanism ensures that earlier incomplete results are revised once more complete data becomes available. Therefore, all false negative matches will be detected, even late, but false positive matches will only be corrected if they are not maximal.

With these strategies in place, the system guarantees that: (i) all emitted matches satisfy the query pattern (soundness), (ii) all matches not involving \texttt{extl}(e) events are eventually detected, even with increased latency, yet, some false positive matches may also be detected (bounded completeness due to extremely late events), and (iii) previously emitted matches are updated when more complete information (i.e. new events) becomes available (repairability).

\paragraph{Example.}
Let us consider the complete in-order event stream \verb|b1 b2 a3 a4 a5 a6 a7 b8 a9 c10| \verb|b11 b12 a13 b14 a15 b16 a17 a18 c19 c20|, with each event having one second gap from the previous. The pattern to be detected is \verb|PATTERN SEQ(A, B+, C) WITHIN 10 secs|. However, due to a \texttt{failure}, the true order of events arrived at the system is \verb|b1 b2 b11 a3 c10 a4 a6 c20 a5 a18 a7 b8| \verb|a17 a9 a13 b14 b16 a15 c19 b12|. 

The end event \texttt{c10} initially triggers match construction but no matches are detected. The same process is followed for the next end event \texttt{c20}, but no matches are detected due to time window constraints. With the arrival of \texttt{b8} as a late event that affects prior matches, the CEP engine is triggered \verb|onDemand|, with MPW containing all events between $[$\texttt{a3} - \texttt{c20}$]$ two times, one with \texttt{c10} and one with \texttt{c20}. The detected matches are:
\begin{center}
\begin{itemize}
    \item matches \texttt{[a3, b8, c10]}, \texttt{[a4, b8, c10]}, \texttt{[a5, b8, c10]}, \texttt{[a6, b8, c10]}, \texttt{[a7, b8, c10]} for \texttt{c10}
    \item no matches for \texttt{c20}
\end{itemize}
\end{center}
Later, late event \texttt{a15} and event \texttt{c19} arrived, so new matches are detected:
\begin{center}
\begin{itemize}
    \item matches \texttt{[a15, b16, c20]} and \texttt{[a13, b14, b16, c20]} are detected due to \texttt{a15} late arrival
    \item matches \texttt{[a15, b16, c19]}, \texttt{[a13, b14, b16, c19]} and \texttt{[a9 b11 b14 b16 c19]} are detected due to \texttt{c19} arrival
\end{itemize}    
\end{center}

However, upon the arrival of late event \texttt{b12}, the match \texttt{[a9, b11, b14, b16, c19]} is invalidated and corrected to \texttt{[a9, b11, b12, b14, b16, c19]} as the former will no longer be maximal. 

Our architecture ensures that as long as events are not \texttt{extl}(e), they are correctly incorporated into the evaluation. Matches are not only sound and (bounded) complete, but also dynamically repairable — enabling reliable, low-latency stream processing under moderate disorder.

\subsection{On replacing SASEXT}
\label{sec:replace}

SASEXT \cite{sasext} is a hybrid CEP execution engine that optimizes pattern detection, especially for cases involving Kleene operators. Unlike traditional NFA-based approaches that construct all possible matches incrementally and store partial results in memory, SASEXT operates in a lazy manner and leverages a hash-based index structure to reduce memory consumption and improve performance. Then, the shared Treeset Structure $STS$ further enhances performance by allowing event sharing across multiple queries, ensuring that common event sequences do not lead to redundant storage and computations.

SASEXT employs a recursive match detection function, starting from an end event $e$ and working backward until a start event is reached. For Kleene patterns, this function ensures that maximal matches are detected, minimizing redundant computations while still allowing the reconstruction of all possible matches as a post-processing step. For instance, given an input stream \texttt{A1 A2 B3 A4 B5 B6 C7} and a pattern \texttt{A+ B+ C}, SASEXT detects the maximal matches \texttt{(A1 A2 B3 B5 B6 C7)} and \texttt{(A1 A2 A4 B5 B6 C7)}. 

Replacing the rationale of SASEXT with another CEP engine while retaining our system's architecture is possible, though it may introduce trade-offs in efficiency and performance. Specifically, an alternative CEP engine, such as SASE, can be integrated while leveraging the existing $STS$ and event management framework for out-of-order and duplicate handling. In such a case, the following observations apply:

\begin{itemize}
    \item \texttt{Regarding Memory and Performance Impact:} Traditional NFA-based engines compute and store intermediate matches, leading to higher memory consumption and processing overhead, compared to SASEXT’s optimized event-indexing approach. This is particularly impactful for large time windows and high event rates, as well as scenarios and patterns with Kleene occurrences.
    
    \item \texttt{Regarding Match Computation Strategy:} While the rationale of SASEXT detects maximal matches and may optionally reconstruct sub-matches as a post-processing step, alternative engines may generate all matches incrementally, potentially increasing computational complexity and memory footprint.
    
    \item \texttt{Regarding Integration with STS:} Any replacement CEP engine will still receive events from $STS$, therefore, ensuring compatibility with the $STS$ is crucial.
\end{itemize}

Overall, despite the fact that LimeCEP is loosely coupled with SASEXT, and can benefit from its optimized memory-efficient processing, integrating another CEP engine is viable. However, careful consideration must be given to the match computation model, memory efficiency, and integration with the existing event storage and processing structure.

\section{Duplicate Handling}
\label{sec:duplicates}

Similar to out-of-order or late arrivals, duplicates in incoming data may cause inaccuracies in the results produced by a CEP engine.
In the Kafka ecosystem, fault tolerance and reliability are crucial aspects, with duplicate message handling being a challenge due to potential consumer failures and message re-deliveries. Kafka’s offset tracking ensures reliable message processing by using unique identifiers to differentiate handled and unhandled messages. While auto-commit policy\footnote{The Kafka auto-commit mechanism allows a consumer to commit the offsets of messages automatically.}  simplifies management, it risks reprocessing duplicates, whereas manual acknowledgment provides more control but introduces additional complexity.

In our solution, Kafka’s mechanisms for reliable message delivery are complemented by $STS$. Kafka’s offset system ensures fault tolerance and minimizes data loss during message re-delivery scenarios. Meanwhile, $STS$ enforces event-level deduplication and ordering, since TreeSets inherently maintain elements in a sorted order and feature a built-in mechanism for duplicate detection and removal. If arriving event $e$ already exists in the corresponding TreeSet, it is simply discarded automatically.

In summary, this dual-layer approach ensures robust duplicate handling. Kafka offsets provide reliability and order guarantees at the transport layer, while 
$STS$ adds application-level accuracy. Therefore, altering the messaging system from Kafka to an MQTT-based protocol does not inherently result in the introduction of duplicate events.

\section{Evaluation}
\label{sec:evaluation}

\subsection{Setup, Competitors, Queries and Datasets}
\label{sec:setup}

All experiments were conducted on a single local machine with 32GB RAM, and 3.8GHz CPU with 8 physical cores and 16 threads. Each experiment was repeated 5 times and the average values are presented. We evaluate four CEP engines, namely SASE\cite{sase}, SASEXT\cite{sasext}, FlinkCEP\cite{flink}, and our approach (LimeCEP). For the evaluation of most experiments, three different types of queries were used, simple query \autoref{q:qA}, complex query with one Kleene+ event present \autoref{q:qB}, complex query with more Kleene+ events \autoref{q:qC}. We examined the behavior of the systems under both selection strategies considered, namely STNM and STAM. The plots of all experiments are provided in the github repository\footnote{\url{https://github.com/KyraStyl/LimeCEP/tree/master/experiments}}.

\begin{tiny}

\begin{lstlisting}[label=q:qA,caption=Q-ABC, captionpos=b,frame=tb]
    PATTERN SEQ(A a, B b, C c)
    WITHIN $W$
\end{lstlisting}

\begin{lstlisting}[label=q:qB,caption=Q-AB+C, captionpos=b,frame=tb]
    PATTERN SEQ(A a, B[] b+, C c)
    WITHIN $W$
\end{lstlisting}

\begin{lstlisting}[label=q:qC,caption=Q-A+B+C, captionpos=b,frame=tb]
    PATTERN SEQ(A[] a+, B[] b+, C c)
    WITHIN $W$
\end{lstlisting}

\end{tiny}

To evaluate the behavior of each engine under realistic conditions, we construct a series of synthetic dataset variants, as well as some out-of-order datasets, derived from base in-order ones. 

We begin with two base datasets:
\begin{itemize}
    \item \textbf{MiniGT-InOrder}: a small synthetic dataset of 20 events with fully known ground truth, designed for correctness evaluation, i.e. accuracy, recall, precision.
    \item \textbf{MicroLatency-10K}: a medium-scale dataset of 10,000 events.
\end{itemize}

From each base dataset, we generated the corresponding out-of-order variant by reordering the generated events to simulate different levels of source inconsistency, with different out-of-order probabilities. More specifically, this includes both a partial out-of-order dataset with probability$\sim$20\%, and a full disorder out-of-order dataset with probability$\sim$70\%, ensuring coverage across a spectrum of real-world conditions.

\begin{table}[h]
\centering
\resizebox{\linewidth}{!}{%
\begin{tabular}{|l|l|r|l|c|l|l|}
\hline
\textbf{Dataset} & \textbf{Base} & \textbf{\#Events} & \textbf{Order} & \textbf{G.T.}  & \textbf{Notes} \\
\hline
MiniGT-InOrder     & —                & 20      & I.O. & \checkmark  & Handcrafted, base for OOO variants \\
MiniGT-PartialOOO  & MiniGT-InOrder   & 20      & P.O. & \checkmark & Same events, mild disorder \\
MiniGT-FullOOO     & MiniGT-InOrder   & 20      & F.O. & \checkmark     & Same events, heavy disorder \\
MiniGT-Duplicates & MiniGT-InOrder & 20 & I.O. & & Same events, multiple duplicates \\
MicroLatency-10K   & —                & 10,000  & I.O. &           & Scalable dataset with clean order \\
MicroLatency-OOO   & MicroLatency-10K & 10,000  & F.O.  &          & Latency-focused OOO variant \\
\hline
\end{tabular}
}
\caption{Overview of datasets used for disorder evaluation. G.T.: Ground Truth, I.O.: In-order, P.O.: Partial Out-of-order (probability$\sim$20\%), F.O.: Full Out-of-order (probability$\sim$70\%)}
\label{tab:datasets}
\end{table}

\subsection{Experiments}
\label{sec:accuracy}

\subsubsection{Impact of out-of-order events on result quality}
\label{sec:exp1a}

\begin{figure}[ht]
    \centering
    \begin{subfigure}[b]{0.8\linewidth}
        \centering
        \includegraphics[width=\linewidth]{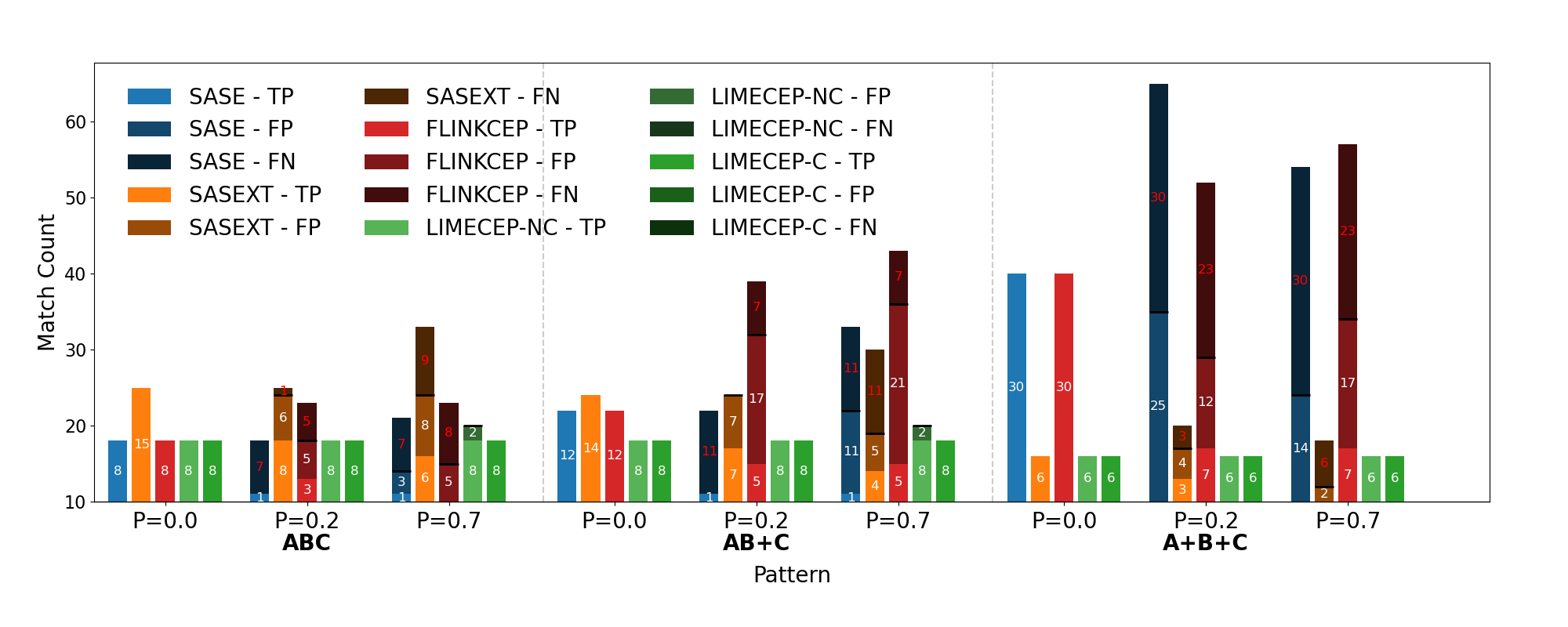}
        \Description{This image shows Accuracy under STNM}
        \caption{Accuracy under STNM}
        \label{fig:accstacked-stnm}
    \end{subfigure}
    \hfill
    \begin{subfigure}[b]{0.8\linewidth}
        \centering
        \includegraphics[width=\linewidth]{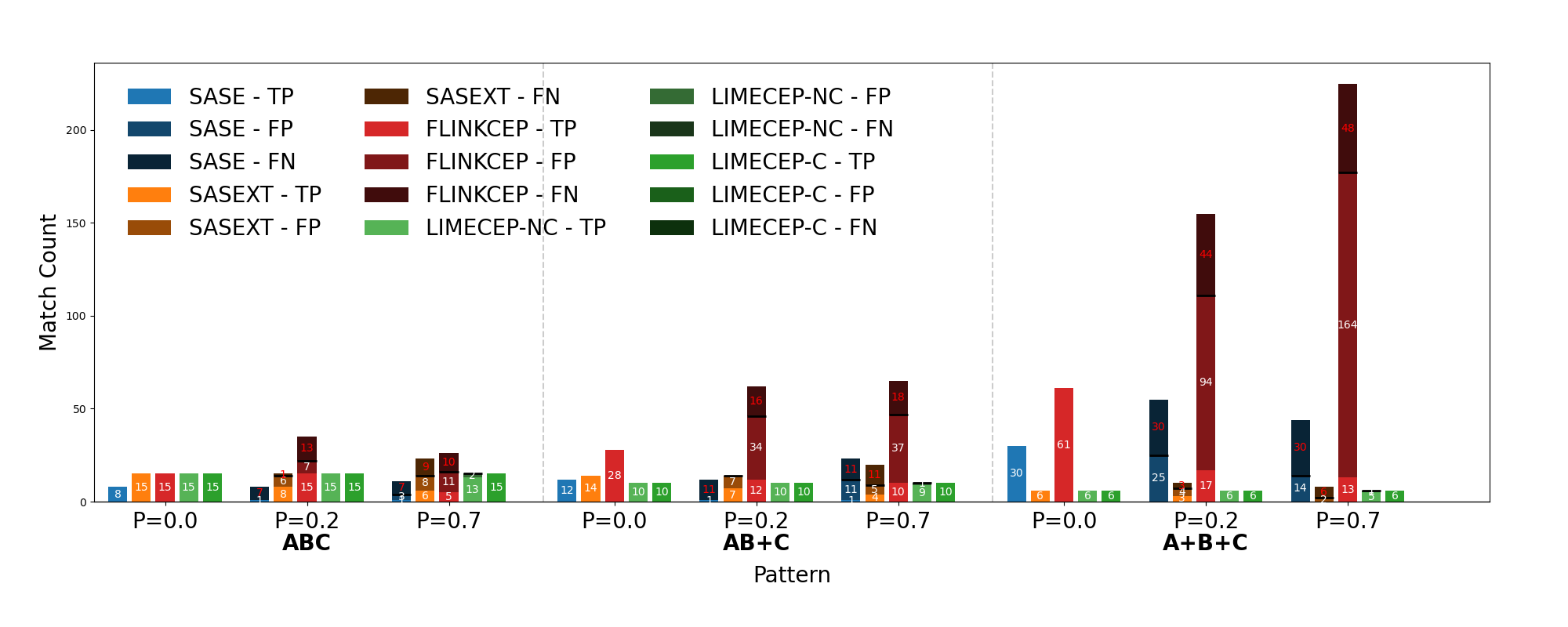}
        \Description{This image shows Accuracy under STAM}
        \caption{Accuracy under STAM}
        \label{fig:accstacked-stam}
    \end{subfigure}
    \caption{Accuracy (TP/FP/FN) across different levels of disorder for STNM and STAM policies.}
    \label{fig:accstacked-combined}
\end{figure}

\begin{figure}[ht]
    \centering
    \begin{subfigure}[b]{0.48\linewidth}
        \centering
        \includegraphics[width=\linewidth]{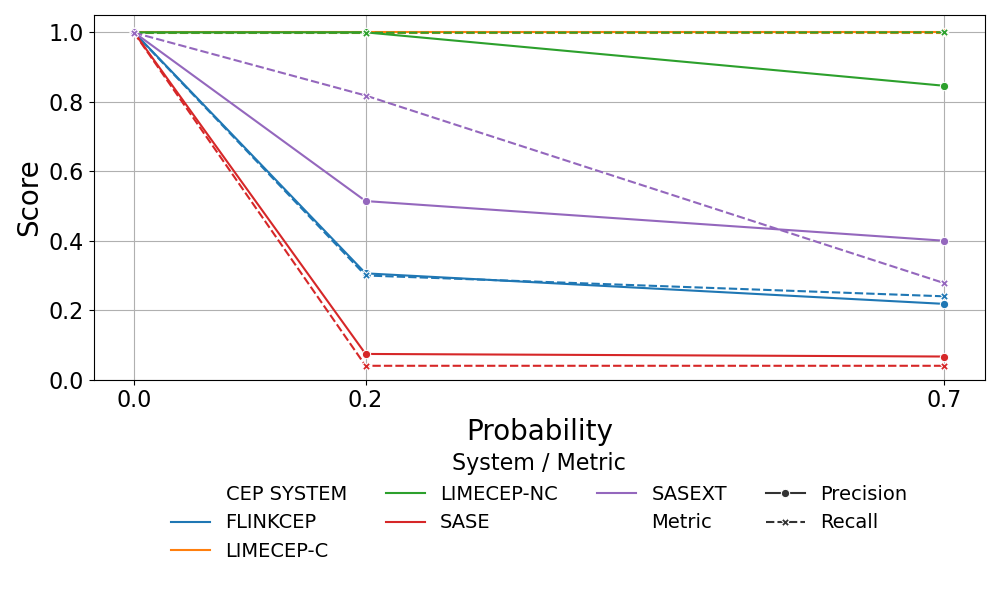}
        \Description{This image shows Recall and Precision on STNM}
        \caption{Recall and Precision on STNM}
        \label{fig:recall-prec-stnm}
    \end{subfigure}
    \hfill
    \begin{subfigure}[b]{0.48\linewidth}
        \centering
        \includegraphics[width=\linewidth]{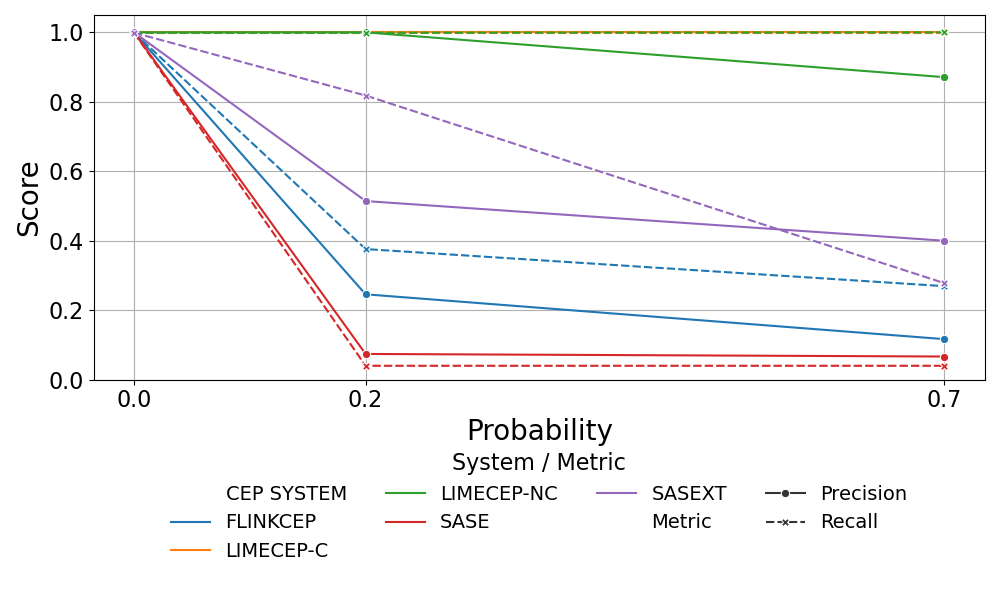}
        \Description{This image shows Recall and Precision on STAM}
        \caption{Recall and Precision on STAM}
        \label{fig:recall-prec-stam}
    \end{subfigure}
    
    \caption{Recall and precision trends under increasing out-of-order probability for STNM and STAM policies.}
    \label{fig:recall-prec-combined}
\end{figure}

In this experiment, we investigate how different CEP engines handle event inconsistency and out-of-order arrivals in streams. We use a small synthetic dataset with ground truth to calculate the recall and precision of each system accurately. Three dataset variants were used, MiniGT-InOrder, MiniGT-PartialOOO, and MiniGT-FullOOO, as described in \autoref{tab:datasets}.

We compare the competitors against two variants of LimeCEP: one with correction enabled when out-of-order events arrive (LimeCEP-C) and one without correction (LimeCEP-NC). For each system, we measure \texttt{precision}, \texttt{recall}, \texttt{true positives} (TP), \texttt{false positives} (FP), and \texttt{false negatives} (FN). As shown in~\autoref{fig:accstacked-stnm} for the STNM policy and in~\autoref{fig:accstacked-stam} for STAM, as the percentage of out-of-order event presence increases, the performance of all systems except LimeCEP-C degrades significantly, especially under heavy OOO conditions. For example, in~\autoref{fig:accstacked-stnm} for $\text{OOO probability} = 0.7$ on pattern \texttt{A+B+C} under STAM policy, FlinkCEP produces 13 FP instead of 61, 164 FP, and 48 FN. SASE detects 0 out of 30 TP matches, 14 FP and 30 FN, while SASEXT operates slightly better, with 0 TP, 2 FP, and 6 FN. In contrast, LimeCEP-C maintains high precision and recall even under heavy out-of-order conditions, consistently producing 6 TP, 0 FP, and 0 FN, while LimeCEP-NC shows a slight degradation, although it still outperforms all other engines, i.e., on \texttt{AB+C} at $\text{OOO probability} = 0.7$, it registers 8 TP, 0 FP, and 2 FN (precision = 1.0, recall = 0.8).

\autoref{fig:recall-prec-stnm} and \autoref{fig:recall-prec-stam} further quantify these trends and illustrate that, while all engines start with perfect accuracy under optimal conditions, i.e., ordered streams ($\text{OOO probability} = 0.0$), the recall and precision of SASE, SASEXT, and FlinkCEP decrease considerably as the probability of late arrivals increases. At $\text{OOO probability} = 0.2$ under STAM, SASE has a precision and recall of around 8-10\%, while FlinkCEP drops to 25-30\% precision and almost 40\% recall. SASEXT has around 50\% precision and around 80\% recall. On the other hand, LimeCEP-C exhibits no degradation in either recall or precision, despite the out-of-order events in the input stream, and LimeCEP-NC exhibits a moderate decline from fully accurate behavior, by achieving around 90\% precision with 100\% recall, only when $\text{OOO probability} = 0.7$. Even though the MiniGT-PartialOOO and MiniGT-FullOOO contain late arrivals and out-of-order events, they do not contain extremely late arrivals (i.e. events that arrive a great amount of time after expected, $extl(e)$ for an event $e$). In this case, even LimeCEP-C will produce false negatives (i.e., miss some matches), although these are still far fewer than the other engines. 


\subsubsection{Impact of duplicate events on result quality}

\begin{figure}
    \centering
    \includegraphics[width=0.75\linewidth]{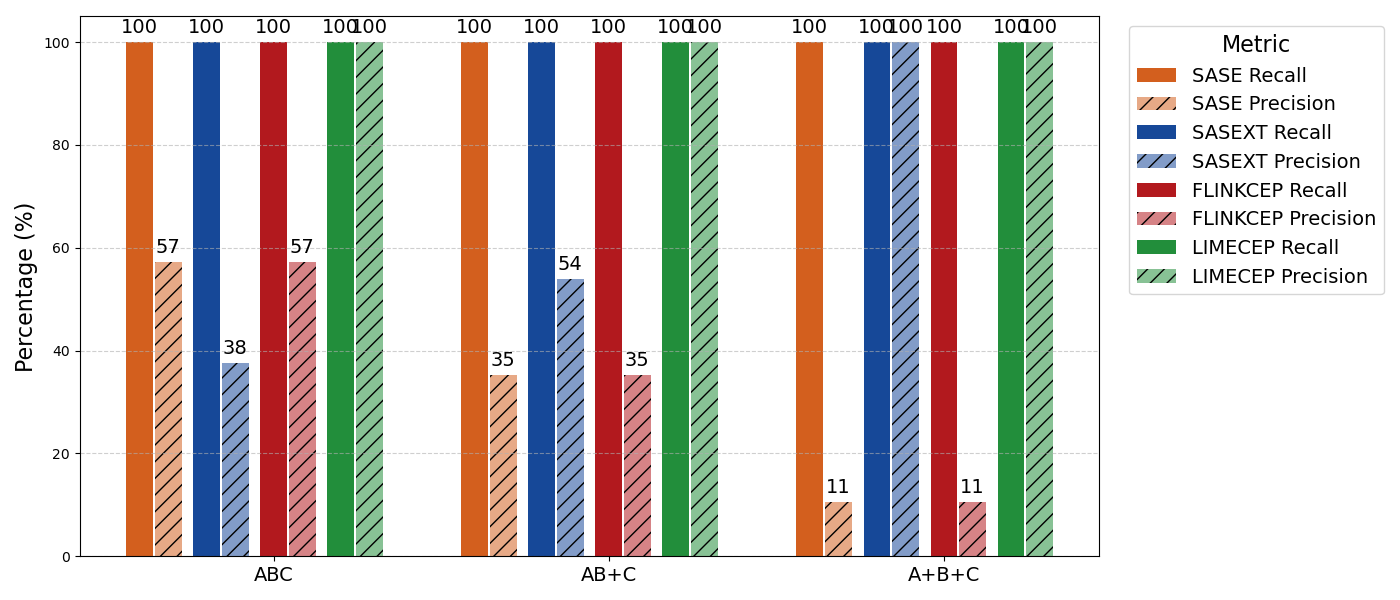}
    \Description{This image shows Recall-Precision STNM vs Duplicate arrivals}
    \caption{Recall-Precision STNM vs Duplicate arrivals}
    \label{fig:duplicates-stnm}
\end{figure}

In this experiment, we evaluate how duplicate events affect the detection quality of different CEP systems under the STNM selection policy. We introduce duplicates for three patterns—\texttt{ABC}, \texttt{AB+C}, and \texttt{A+B+C}—and measure each system’s \texttt{recall} and \texttt{precision} in the presence of these duplicates.

\autoref{fig:duplicates-stnm} depicts that even though all systems maintain 100\% recall across all patterns, i.e., none of the systems provides false negatives, their precision degrades significantly. SASE and FlinkCEP show similar behavior, with their precision dropping sharply, as the complexity of the pattern increases, since each arriving duplicate event may affect and therefore produce multiple matches. More specifically, for the \texttt{ABC} pattern, SASE and FlinkCEP detect additional 6 FP matches (detecting 14 matches overall with 8 TP), causing precision to drop to 57\%. SASEXT performs worse in this pattern by detecting 25 FP with precision dropping to 38\%, however, it performs better in more complex patterns (i.e. \texttt{AB+C} and \texttt{A+B+C}) since detecting maximal matches decreases the number of matches produced per each duplicate event, therefore, showing improved precision, by detecting only 10 out of 26 FP in the first case, and none in the second.

LimeCEP, on the other hand, consistently maintains high precision on pattern detection, by eliminating duplicate output across all patterns, detecting zero FP matches in every case. This result confirms that LimeCEP can handle duplicate arrivals efficiently, avoiding false positives entirely.

\subsubsection{Parameter sensitivity of LimeCEP}
\label{sec:exp1b}

\begin{figure}[ht]
    \centering
    \begin{subfigure}[b]{0.6\linewidth}
        \centering
        \includegraphics[width=\linewidth]{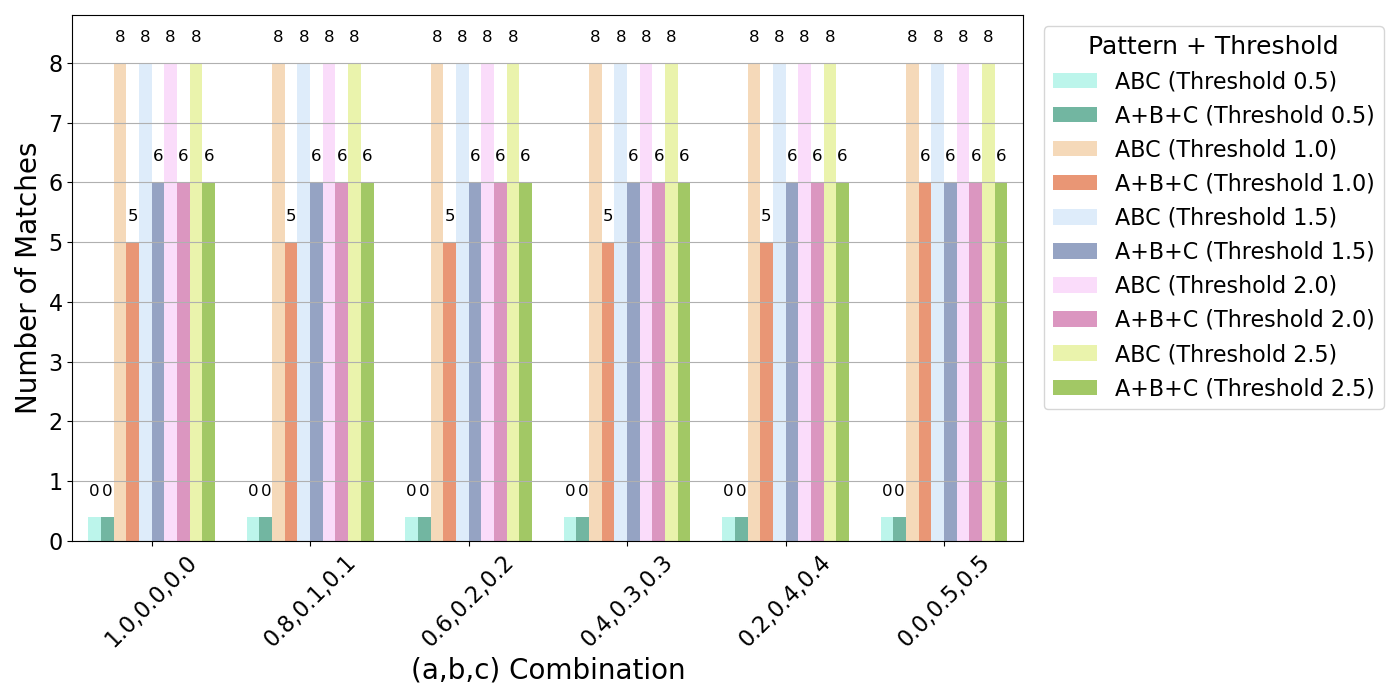}
        \Description{This image shows Effect of the lateness threshold under STNM}
        \caption{STNM}
        \label{fig:exp1b-stnm}
    \end{subfigure}
    \hfill
    \begin{subfigure}[b]{0.6\linewidth}
        \centering
        \includegraphics[width=\linewidth]{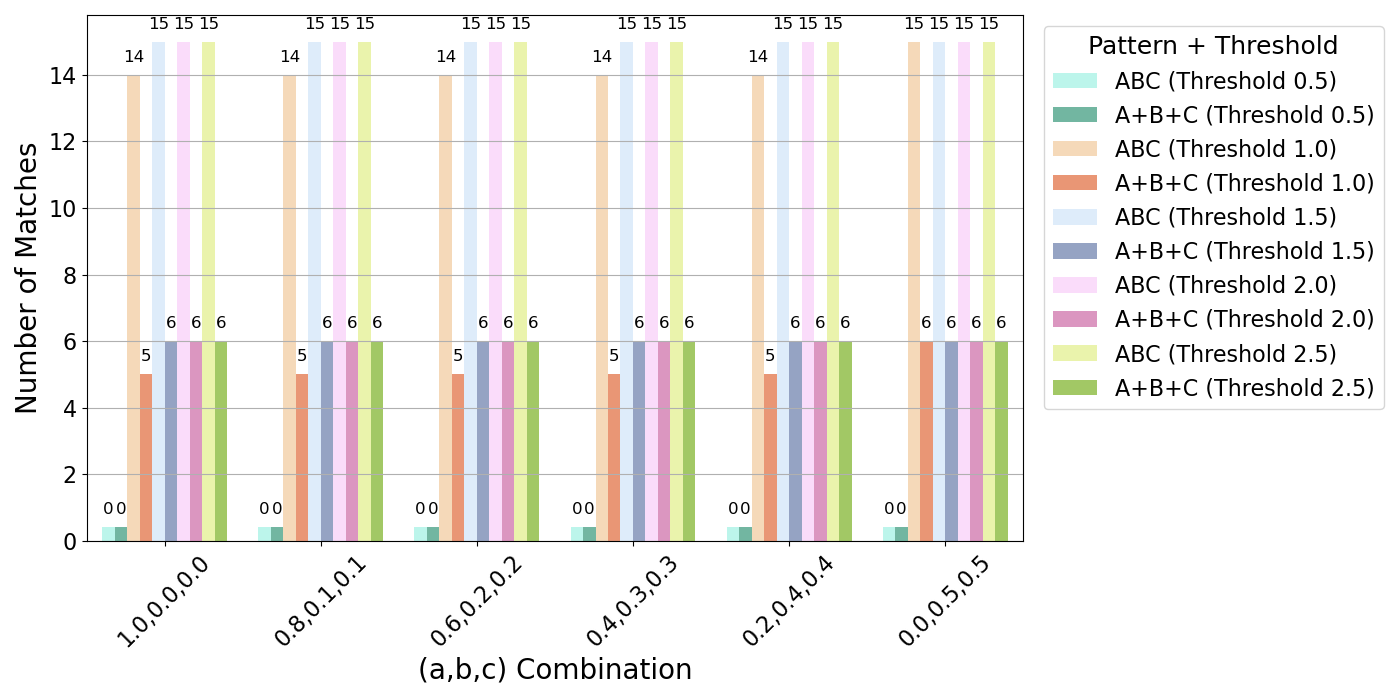}
        \Description{This image shows Effect of the lateness threshold under STAM}
        \caption{STAM}
        \label{fig:exp1b-stam}
    \end{subfigure}
    \caption{Effect of the lateness threshold ($\theta_{s_{et}}$) and scoring weights $(a,b,c)$ on detection accuracy under high disorder ($\text{OOO probability} = 0.7$).}
    \label{fig:exp1b-combined}
\end{figure}

This experiment evaluates the effect of two core LimeCEP parameters on accuracy:
\begin{itemize}
    \item Late event threshold $\theta_{s_{et}}$, which defines the maximum tolerated delay for out-of-order events per event source.
    \item Scoring weights $(a,b,c)$ used in the $ooo(e)$ function.
\end{itemize}

We evaluated two patterns, \texttt{ABC}(\autoref{q:qA}) and \texttt{A+B+C}(\autoref{q:qC}), under a high disorder probability ($\text{OOO probability} = 0.7$) and systematically vary $\theta_s$ and the $(a,b,c)$ combinations to assess their individual and joint impact on result quality. 

~\autoref{fig:exp1b-combined} depicts that when $\theta_{s_{et}}$ is low ($\leq 0.5$), all configurations fail to detect any matches with recall dropping to 0, in both cases, STNM and STAM. This occurs because delayed events are discarded before they can contribute to pattern matching.
At $\theta_{s_{et}} = 1.0$, for the STNM policy recall increases to 100\%. However, some of the events arrived in the system with a high out-of-order score, which could affect the detected matches in other patterns. This is more evident under the STAM policy, where recall increases to approximately 83.3\% for STNM (5 out of 6 for A+B+C on STNM policy) and 93.3\% for STAM (14 out of 15 correct matches on STAM are detected), regardless of the $(a,b,c)$ weights. This happens due to late arrivals with high scores, which did not fully contribute to the matches produced, i.e., an event should trigger match correction, but did not, due to the high score.
For $\theta_{s_{et}} \geq 1.5$, all combinations achieve recall=100\% and precision=100\%. This proves that LimeCEP can successfully detect all matches across different user-defined values of its factors, indicating a robust behavior.

In addition to the lateness threshold, different combinations of $(a,b,c)$ were tested on whether and how they affect the accuracy achieved by LimeCEP in cases of late arrivals. As depicted in ~\autoref{fig:exp1b-combined}, factors of out-of-orderness score seem to have little effect on recall or precision and no effect once the threshold is tolerant enough. For the case of $\theta_{s_{et}} = 1.5$, the best combination seems to be the one that does not count at all the time difference of an arriving event for the ooo\_score. Even in more extreme cases (e.g., $a=1.0$, $b=c=0.0$ or uniform $a=b=c=0.3$) same accuracy is achieved for $\theta_{s_{et}} \geq 1.5$. This validates that LimeCEP is robust to its internal configuration.

\subsubsection{Maximum Detection Latency}

\begin{figure}[ht]
    \centering
    \begin{subfigure}[b]{0.48\linewidth}
        \centering
        \includegraphics[width=\linewidth]{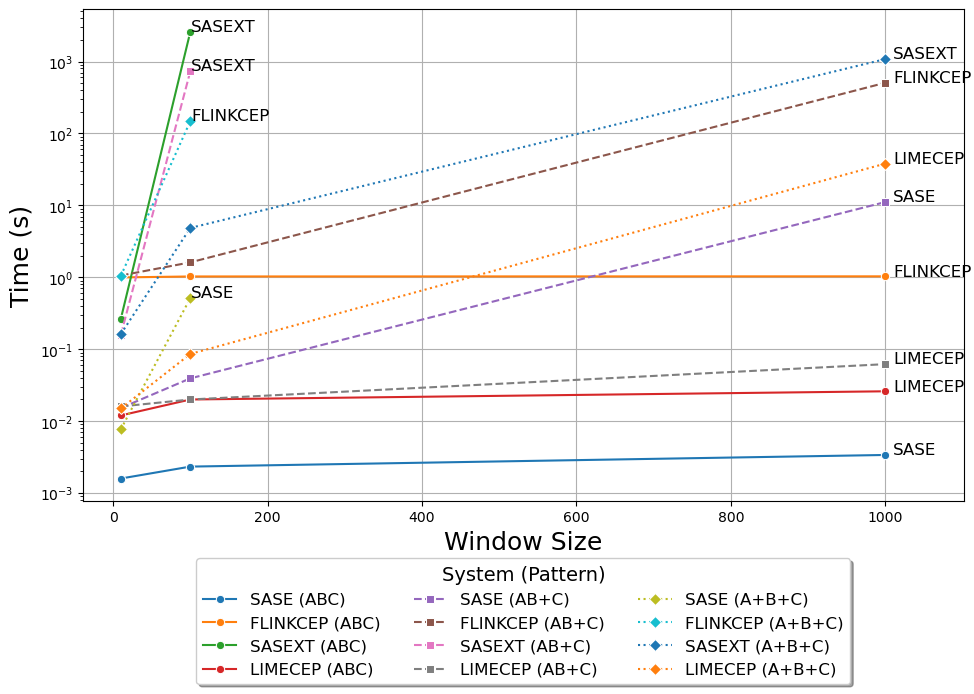}
        \Description{This image shows the Maximum latency of systems under STNM selection policy.}
        \caption{STNM}
        \label{fig:exp2-stnm}
    \end{subfigure}
    \hfill
    \begin{subfigure}[b]{0.48\linewidth}
        \centering
        \includegraphics[width=\linewidth]{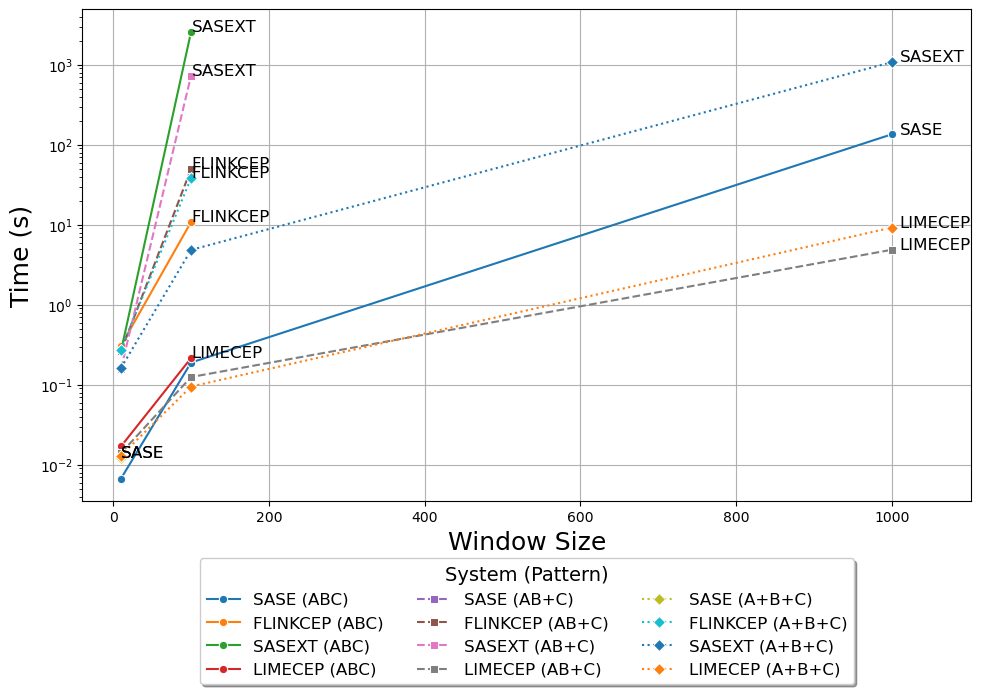}
        \Description{This image shows the Maximum latency of systems under STAM selection policy.}
        \caption{STAM}
        \label{fig:exp2-stam}
    \end{subfigure}
    \caption{Maximum latency of systems under STNM and STAM selection policies.}
    \label{fig:exp2-combined}
\end{figure}

In this experiment, we measure the maximum latency required by each CEP system to detect a complete pattern match. We vary the window size and the pattern complexity across three configurations: \texttt{ABC} (\autoref{q:qA}), \texttt{AB+C} (\autoref{q:qB}), and \texttt{A+B+C} (\autoref{q:qC}). Latency is measured in nanoseconds (ns), and results are reported separately for STNM and STAM policies in \autoref{fig:exp2-combined}.

Under the STNM policy, SASE in \texttt{ABC} pattern achieves the lowest latency overall, staying below $10^7$ ns (under 10 milliseconds). Next is LimeCEP (ABC), with latency around $10^8$ ns ($\approx$0.1 seconds), which is two orders of magnitude faster than FlinkCEP (ABC), which reaches $10^{10}$ ns ($\approx$10 seconds). We can observe that as the pattern becomes more complex, SASE fails to retain the latency low, in some cases, e.g. in \texttt{A+B+C} with window 100 or 1000, it even fails to run completely.

On the other hand, LimeCEP maintains stable latency for both \texttt{ABC} and \texttt{AB+C} patterns. More specifically, for pattern \texttt{AB+C} it retains latency around $10^8$ ns, while FlinkCEP grows to $10^{12}$ ns ($\approx$1,000 seconds or $\approx$16 minutes), resulting in a difference of four orders of magnitude between the two. In the \texttt{A+B+C} pattern, LimeCEP ranges between $10^7$ and $10^{11}$ ns (up to $\approx$100 seconds depending on the window), while FlinkCEP starts around $10^9$ ns (1 second) and eventually fails to complete due to excessive memory consumption, effectively producing unbounded latency in some cases.

Under the STAM policy, most of the cases, especially the ones with larger windows, did not run at all, either due to memory exceeding or due to time exceeding a threshold of two hours. LimeCEP achieves the smallest latency in almost all cases, regardless of the pattern or window size, except for the case of \texttt{ABC} with a window equal to 1000, which failed due to memory. That happened due to the correction feature of LimeCEP, which retains multiple indices that assist in the correction or invalidation of detected matches. In case this feature is not enabled, all cases run with lower resource consumption and also lower latency detection. 

Across configurations, LimeCEP's latency remains within $10^7$ to $10^9$ ns (from 10 milliseconds to 1 second), while other systems, especially SASEXT and FlinkCEP, still suffer latency spikes under complex patterns and large windows. Notably, SASEXT reaches up to $10^{12}$ ns under the A+B+C pattern even in STAM policy. Overall, LimeCEP demonstrates up to 6 orders of magnitude lower latency, compared to other CEP engines.

\subsubsection{Resource Consumption}

\begin{figure}[ht]
    \centering
    \begin{subfigure}[b]{0.48\linewidth}
        \centering
        \includegraphics[width=\linewidth]{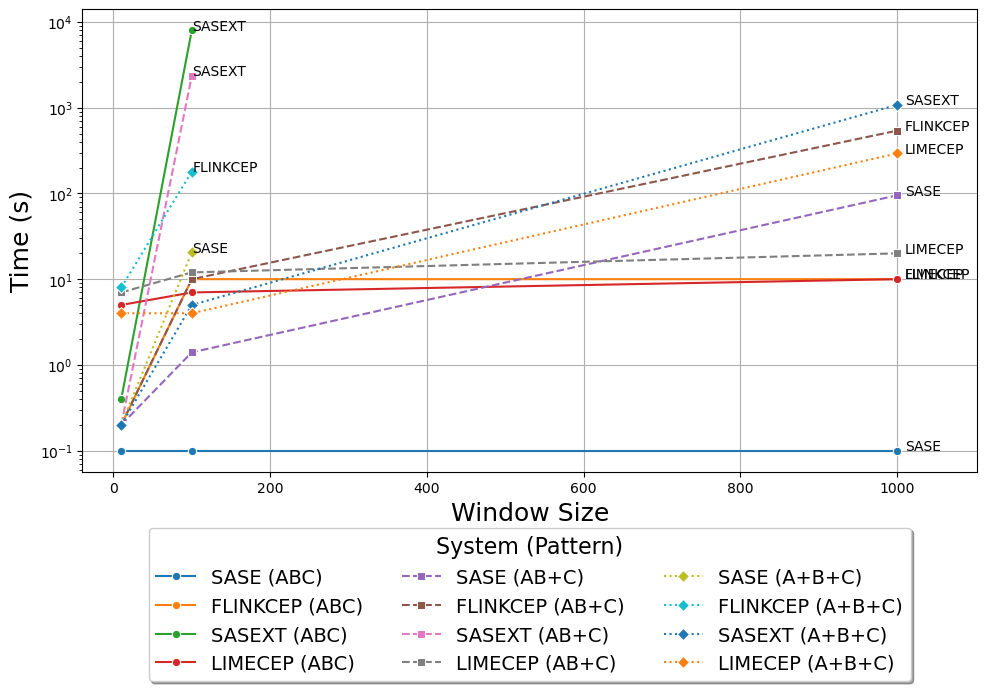}
        \Description{This image shows the Execution time comparison under STNM selection policy.}
        \caption{STNM}
        \label{fig:exp3-exec-time-stnm}
    \end{subfigure}
    \hfill
    \begin{subfigure}[b]{0.48\linewidth}
        \centering
        \includegraphics[width=\linewidth]{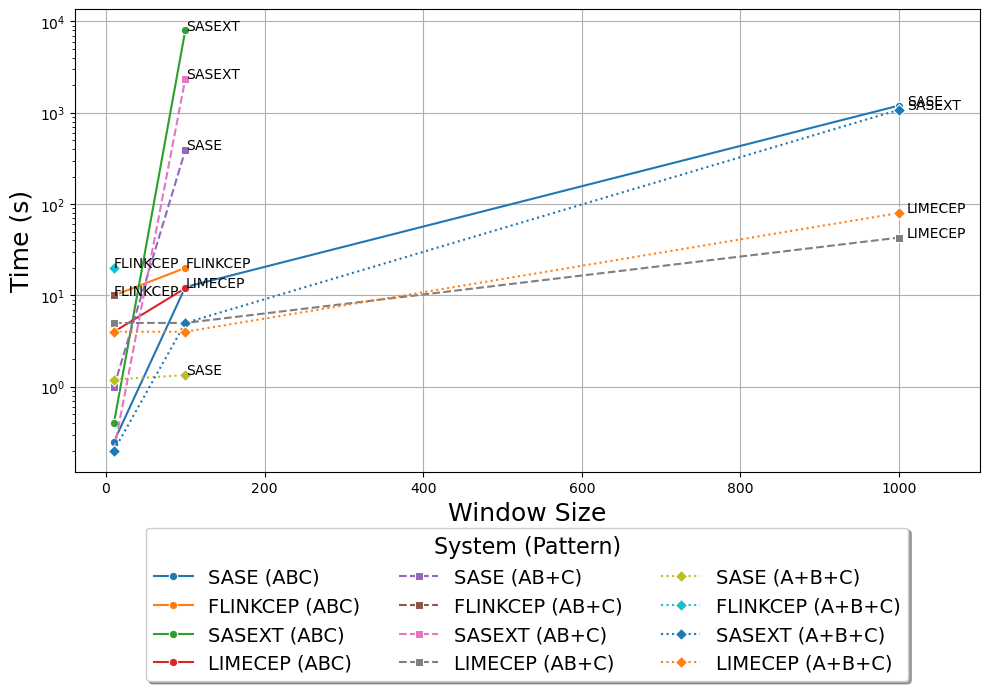}
        \Description{This image shows the Execution time comparison under STAM selection policy.}
        \caption{STAM}
        \label{fig:exp3-exec-time-stam}
    \end{subfigure}
    \caption{Execution time comparison under STNM and STAM selection policies.}
    \label{fig:exp3-exec-time-combined}
\end{figure}

\begin{figure}[ht]
    \centering
    \begin{subfigure}[b]{0.48\linewidth}
        \centering
        \includegraphics[width=\linewidth]{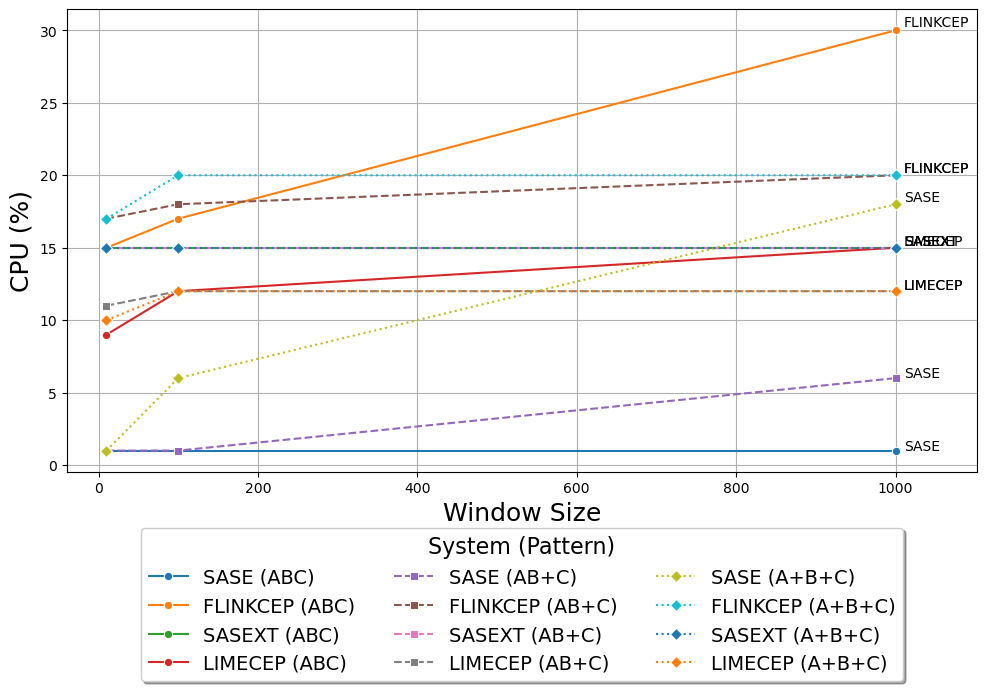}
        \Description{This image shows CPU usage under STNM selection policy.}
        \caption{STNM}
        \label{fig:exp3-cpu-stnm}
    \end{subfigure}
    \hfill
    \begin{subfigure}[b]{0.48\linewidth}
        \centering
        \includegraphics[width=\linewidth]{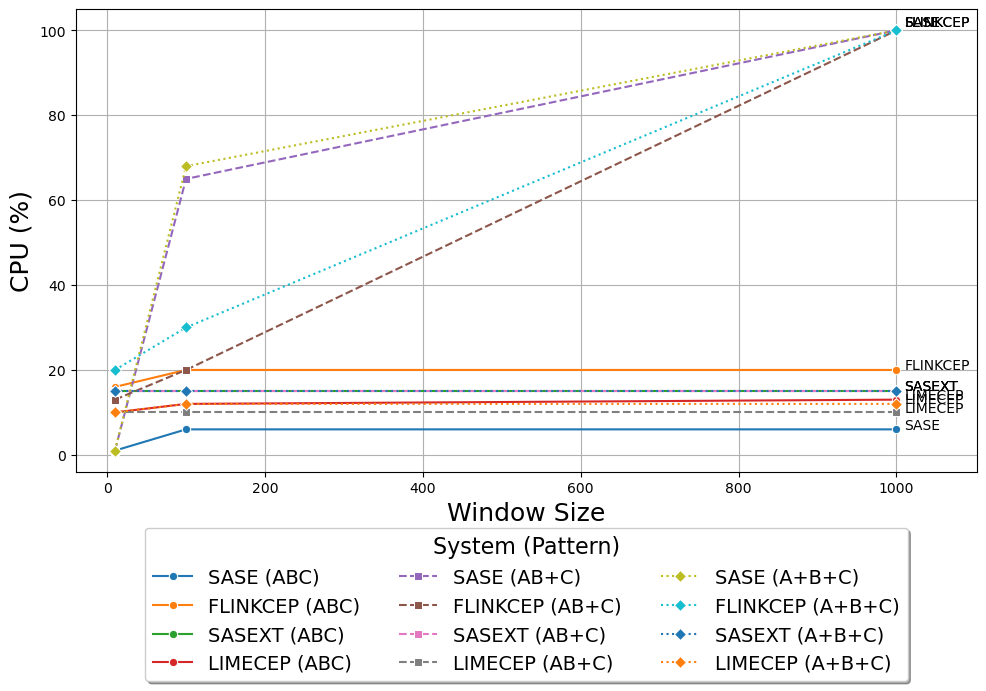}
        \Description{This image shows CPU usage under STAM selection policy.}
        \caption{STAM}
        \label{fig:exp3-cpu-stam}
    \end{subfigure}
    \caption{CPU usage under STNM and STAM selection policies.}
    \label{fig:exp3-cpu-combined}
\end{figure}

\begin{figure}[ht]
    \centering
    \begin{subfigure}[b]{1.0\linewidth}
        \centering
        \includegraphics[width=\linewidth]{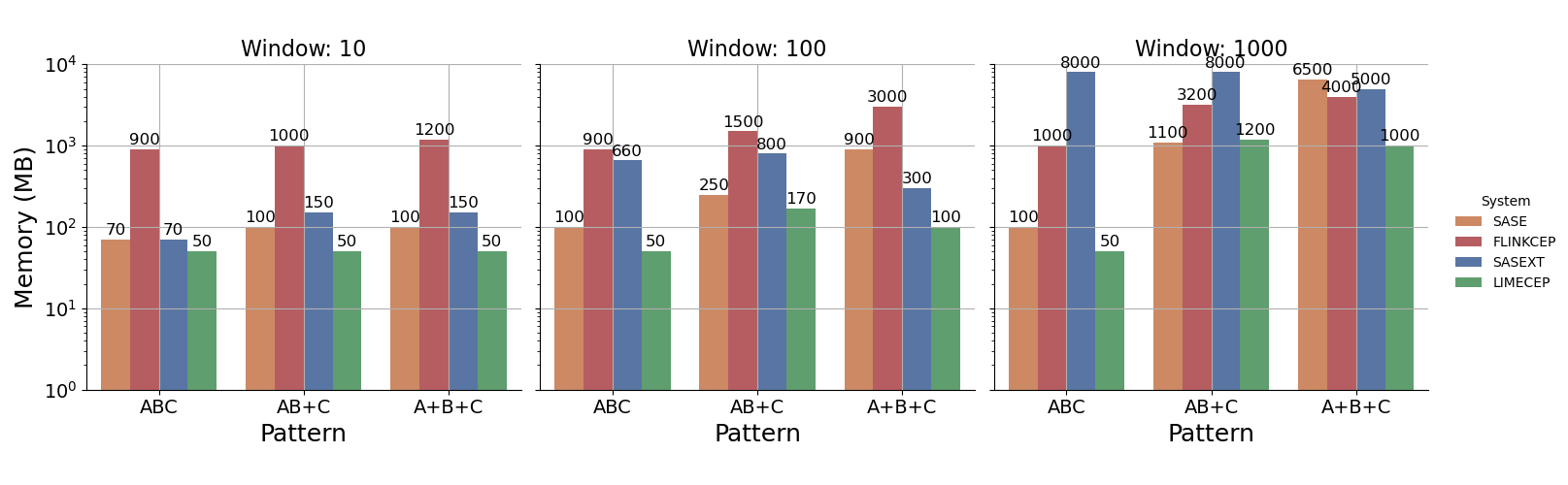}
        \Description{This image shows Memory usage comparison under STNM selection policy.}
        \caption{STNM}
        \label{fig:exp3-mem-stnm}
    \end{subfigure}
    \hfill
    \begin{subfigure}[b]{1.0\linewidth}
        \centering
        \includegraphics[width=\linewidth]{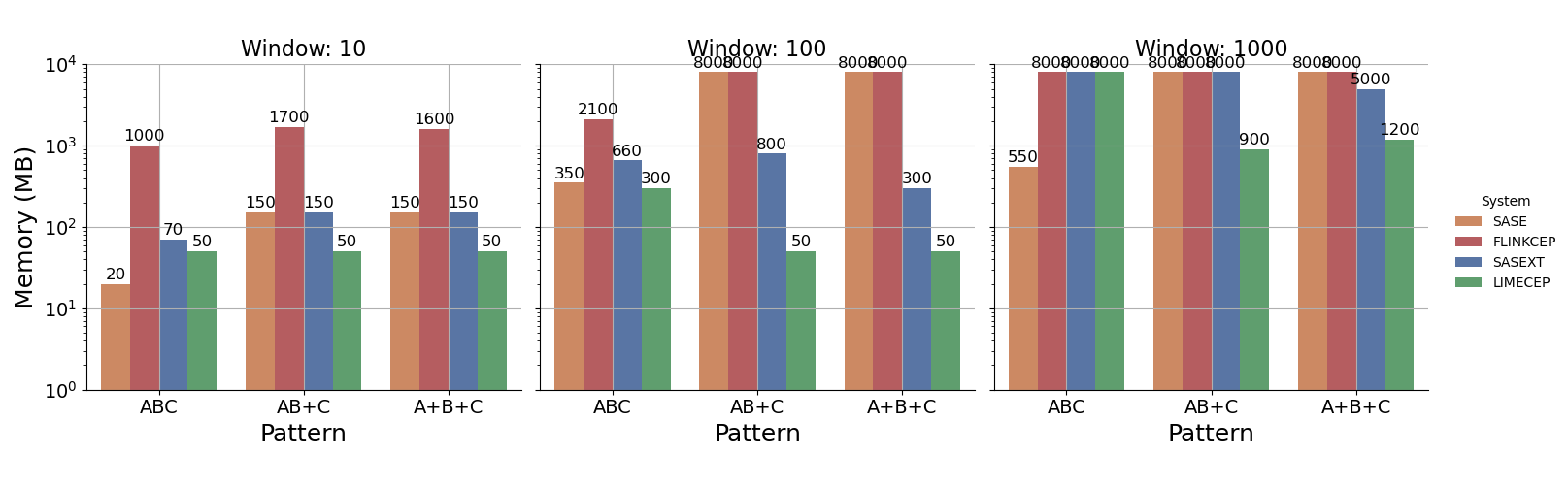}
        \Description{This image shows Memory usage comparison under STAM selection policy.}
        \caption{STAM}
        \label{fig:exp3-mem-stam}
    \end{subfigure}
    \caption{Memory usage comparison under STNM and STAM selection policies.}
    \label{fig:exp3-mem-combined}
\end{figure}

In this experiment, we evaluate the resource usage of the CEP systems with respect to three main metrics: CPU utilization (\autoref{fig:exp3-cpu-combined}), total execution time (\autoref{fig:exp3-exec-time-combined}), and memory consumption (\autoref{fig:exp3-mem-combined}). The systems are tested under varying window sizes and pattern (ABC, AB+C, A+B+C), using both STNM and STAM selection policies.

\textbf{CPU.} Under the STNM policy, CPU consumption increases significantly for systems like SASE and SASEXT as both window size and pattern complexity grow. For example, SASE and SASEXT reach close to 100\% CPU usage under large windows and relaxed patterns such as A+B+C. In contrast, LimeCEP maintains consistently low CPU usage, remaining below 15\% across all configurations. 
Under STAM, FlinkCEP, and SASE show linear growth with window size, with the cases of window size equal to 1000 failing to run completely. LimeCEP again exhibits the lowest CPU usage overall, typically staying between 10–15\%, regardless of pattern structure.

\textbf{Execution time.} Execution time results further emphasize the efficiency of LimeCEP. Under STNM, both SASE and SASEXT scale poorly, reaching up to 10,000 seconds (nearly 3 hours) for large windows and complex patterns. FlinkCEP also shows significant growth, particularly with relaxed patterns.
In contrast, LimeCEP maintains low total execution times, ranging from under 10 seconds for \texttt{ABC}, to under 300 seconds even in the most demanding \texttt{A+B+C} configuration. For large windows, it executes 10× to 1,000× faster than SASEXT or FlinkCEP, depending on the pattern and policy. 
Under the STAM policy, LimeCEP again outperforms the other engines in most of the patterns and windows, achieving sub-10-second runs for most of the cases, and under 100 seconds even for the most demanding case of \texttt{A+B+C} pattern within window 1000.

\textbf{Memory consumption.} Memory usage was measured per pattern across three window sizes (10, 100, and 1000) in log scale. For small windows, all engines are efficient, but differences appear as the window grows.

SASE has the lowest memory usage for small windows but grows exponentially with window size and pattern complexity. Under both STNM and STAM, FlinkCEP and SASEXT exceed 8,000 MB (8 GB) for large windows in \texttt{AB+C} and \texttt{A+B+C} patterns, while SASE also grows up to 6.5 GB of memory. In contrast, LimeCEP remains highly memory-efficient, using only 1.2 GB for the most complex case, and as low as 50–300 MB for smaller or simpler setups.

Across all evaluated configurations, LimeCEP consistently consumes one to two orders of magnitude less memory than the other CEP systems. For simple patterns such as ABC at large window sizes, LimeCEP requires only 50 MB, whereas SASE reaches 550 MB (a difference of one order of magnitude), and both SASEXT and FlinkCEP exceed 8000 MB (more than two orders of magnitude higher). For more relaxed patterns like AB+C and A+B+C, the difference remains substantial but not over one order of magnitude (from 1200MB to 5000 or 8000 MB).

Overall, LimeCEP proves to be the most resource-efficient system across all metrics evaluated. It consistently achieves low CPU utilization, fast execution times, and controlled memory usage, even when processing complex patterns over large windows.

\subsubsection{Memory under multiple pattern detection}
\label{sec:exp5-multipattern}

\begin{figure}
    \centering
    \includegraphics[width=1.0\linewidth]{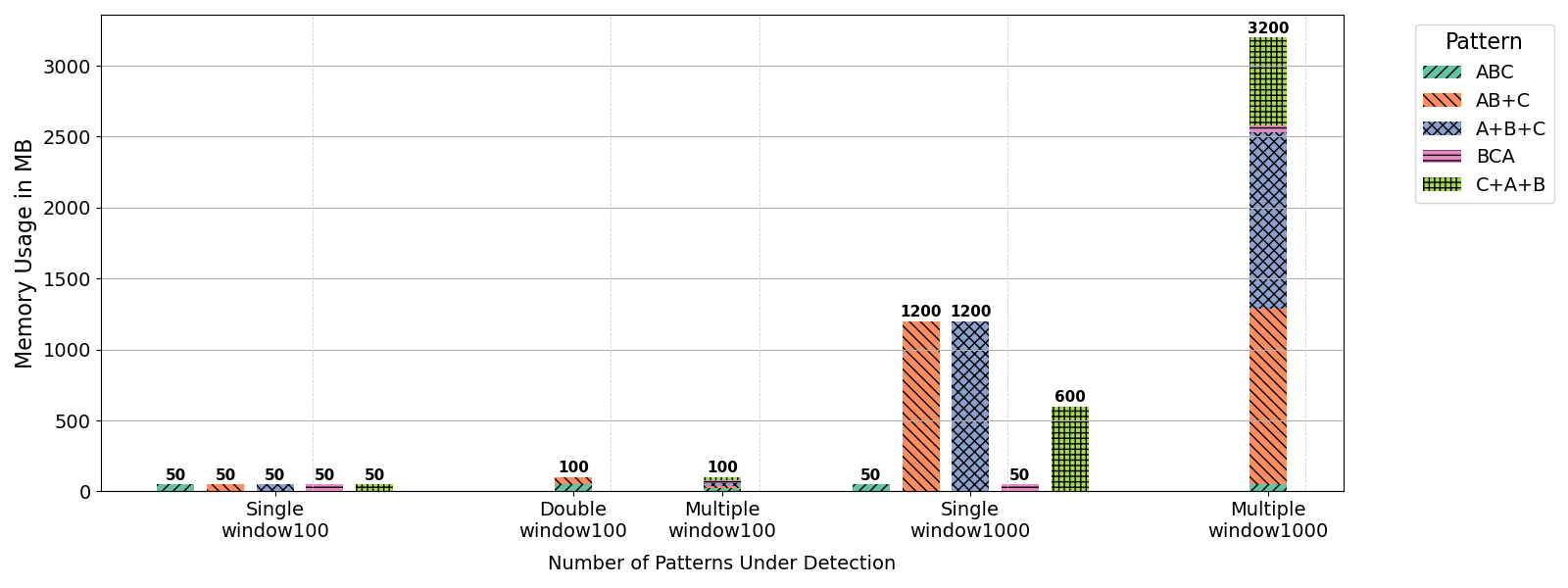}
    \Description{This image shows Multiple pattern execution}
    \caption{Multiple patterns}
    \label{fig:multipattern}
\end{figure}

In this experiment, we assess how LimeCEP scales with respect to memory usage when detecting multiple patterns simultaneously. We measure total memory consumption under moderate (100) and large (1000) window sizes, under different configurations: (i) detection of individual patterns, (ii) detection of a pair of patterns, and (iii) detection of five patterns concurrently. The results are presented in \autoref{fig:multipattern}

At window size 100, LimeCEP demonstrates sublinear memory growth with respect to the number of patterns under detection. Despite processing up to five distinct patterns, total memory usage remains capped at 100MB, indicating that LimeCEP reuses internal structures and avoids redundant state maintenance. This design results in minimal memory overhead even in multi-pattern workloads.
At window size 1000, LimeCEP’s memory usage increases depending on the complexity of each pattern. Simple patterns like ABC and BCA still use only 50MB, while relaxed patterns such as AB+C and A+B+C require up to 1200MB each. When all five patterns are running at once, total memory usage reaches 3200MB. This shows that memory grows with pattern complexity, but not in a wasteful or exponential way. LimeCEP handles multiple complex patterns efficiently, without using more memory than needed.

\section{Conclusions}
\label{sec:concl}

In this work, we introduced LimeCEP, a novel complex event processing engine designed to effectively manage the challenges of real-world data streams, including out-of-order, late, and duplicate event arrivals, using constrained resources. Our hybrid approach combines lazy processing with both buffering and speculative processing along with the ability to correct results, and also supports multi-pattern detection using shared structures, and maximal match detection to limit the required processing power and memory consumption, enabling correct and timely results.

Through our extensive evaluation, LimeCEP demonstrates a clear advantage over existing CEP systems, such as the SASE prototype, SASEXT (an extension of SASE), and the FlinkCEP library of the Apache Foundation. It maintains near-perfect precision and recall even under a high-disorder input stream, delivers up to six orders of magnitude lower latency, and uses significantly less memory and CPU power. LimeCEP’s ability to support relaxed and non-deterministic semantics even in demanding scenarios where other CEP engines such as SASE and FlinkCEP fail, combined with its resource-aware design and lazy evaluation, makes it particularly suitable for non-cloud environments. By supporting multiple pattern detection and handling inconsistencies in input data, it can be used in many real-world scenarios. 

In the future, to further enhance LimeCEP's functionality, we aim to extend it with mechanisms to account for potentially missing events. Finally, we envision augmenting LimeCEP with automated pattern discovery capabilities, allowing it to identify meaningful patterns beyond the ones explicitly defined by the user.

\bibliographystyle{ACM-Reference-Format}
\bibliography{references}

\end{document}